\documentclass[11pt]{article}
\usepackage{amssymb,amsmath,amsfonts,mathrsfs,enumerate,wrapfig}
\usepackage{graphicx,rotate,multicol,epsfig}
\usepackage{tikz}
\usepackage[utf8]{inputenc}
\usepackage{tabulary}
\usepackage{bm}

\usepackage[newcommands]{ragged2e}
\usepackage{cite}
\usepackage{pdfpages}
\usepackage[margin = 1 in]{geometry}
\usepackage{graphicx}
\usepackage{slashed}
\usepackage{amsmath}

\usepackage[colorlinks=true,
            linkcolor=red,
            urlcolor=blue,
            citecolor=blue]{hyperref}
\usepackage{pstricks} 
\usepackage{multirow} 
\usepackage{slashed}

\usepackage{graphicx}
\usepackage{slashed}
\usepackage{color}
\usepackage{tikz}

\def \order(#1){{\cal O} \left(#1 \right)}
\long\def\rpl#1!!#2!!{\textcolor{red}{#1} \textcolor{blue}{#2}}

\evensidemargin=\oddsidemargin
\allowdisplaybreaks

\def\bar {\overline}

\def\be {\begin{equation}}
\def\ee {\end{equation}}
\def\beq {\begin{equation}}
\def\eeq {\end{equation}}
\def\bea {\begin{eqnarray}}
\def\eea {\end{eqnarray}}

\def\beq{\begin{equation}}
\def\eeq{\end{equation}}
\def\barr{\begin{array}}
\def\earr{\end{array}}

\def\opcit(#1){ {\em op. cit.}, #1}

\usepackage[normalem]{ulem}

\definecolor{darkgreen}{cmyk}{1,0,1,0.4}
\definecolor{pink}{cmyk}{0.4,1,0.3,0}

\begin{document}

\renewcommand*{\thefootnote}{\fnsymbol{footnote}}

\begin{center}
 {\Large\bf{Cut and compute: Quick cascades with multiple amplitudes}}\\[5mm]
\vspace{0.5cm}
{\bf Joydeep Chakrabortty} $^a$\footnote{joydeep@iitk.ac.in},
{\bf Anirban Kundu} $^b$\footnote{anirban.kundu.cu@gmail.com}, 
{\bf Rinku Maji}$^a$\footnote{mrinku@iitk.ac.in},  
{\bf Tripurari Srivastava} $^a$\footnote{tripurar@iitk.ac.in}

{\small
$^a$ {Department of Physics, Indian Institute of Technology, Kanpur 208016, India}\\[1mm]
$^b$ {Department of Physics, University of Calcutta, \\
92 Acharya Prafulla Chandra Road, Kolkata 700009, India}
}

\end{center}
\begin{abstract}
In an earlier paper \cite{Chakrabortty:2016idh}, we have proposed a novel method to compute the decay width 
for a general $1\to n$ cascade decay where the propagators are off-shell and may be of different spins. 
Here, we extend our algorithm to accommodate those decays that are mediated by more than one such 
cascades. This generalizes our prescription and widens its applicability. We compute the  three- 
and four-body toy decay chains where identical final states appear through different cascades. Here, 
we also provide the algorithm to calculate the interference terms. For four-body decays we discuss 
both symmetric and asymmetric cascades, providing the expressions for the detailed phase space 
structure in each case. We find that the results obtained with this algorithm have a very impressive 
agreement with those from standard softwares using a sophisticated Monte Carlo based phase space integration.

\end{abstract}

\date{\today} 


\setcounter{footnote}{0}
\renewcommand*{\thefootnote}{\arabic{footnote}}

\section{Introduction}

Field theoretic calculation of the decay width for $1\to n$ processes gets complicated as $n$ increases,
unless the Feynman diagram can be decomposed into smaller parts by cutting through the on-shell lines. 
While it is not allowed to cut the off-shell lines, one may ask whether it is possible to find some algorithm where 
the $1\to n$ diagrams can be effectively decomposed in a number of $1\to 2$ subdiagrams with possible 
off-shell incoming and outgoing legs.    

We have addressed this issue and come up with such an algorithm in our earlier paper 
\cite{Chakrabortty:2016idh}, also tested against the standard numerical packages for simulating the 
multibody phase space. We have defined a modified $|\mathcal{M}|^2$ which is a matrix 
spanned over the bases of quantum numbers being carried by the off-shell propagators.
We have shown that one can decompose the full cascade into several $1\to 2$-body decays, which may 
contain off-shell particles as incoming or outgoing legs. Thus, one
needs to compute only the $1\to 2$-body decays, which is a standard textbook exercise. 
Then, following our proposal, all those 2-body decays can be combined, leading to the result for the 
full $1\to n$-body decay. In essence, one has to remember two points: (i) In the calculation of the 
matrix element squared, $p^2$ for an off-shell field has to be replaced by its invariant mass squared and 
not its physical mass squared, and then one has to integrate over the entire range of the invariant mass, and 
(ii) For the spin sum, the completeness relation should contain the physical mass, and not the invariant 
mass.

In this paper, we extend our earlier proposal to include decays that proceed through more than one cascades
leading to an identical final state. This necessitates the inclusion of interference diagrams, and we show
how to take these diagrams systematically into account, even when the off-shell propagators have 
different spin (and other quantum numbers). 
For the help of the reader, we also compute some explicit examples of $1\to 3$-body decays, and 
compare the results obtained using the software {\tt CalcHEP} vis-a-vis our proposal. The results are 
in excellent agreement, but of course an implementation of our algorithm will be faster to execute. 
We have also shown how to deal with $1\to 4$ decays, with topologically different cascades. 

\begin{figure}[h!]
\begin{center}
\includegraphics[trim={0 9cm 0 3cm},clip,width=15 cm,height=9.0cm]{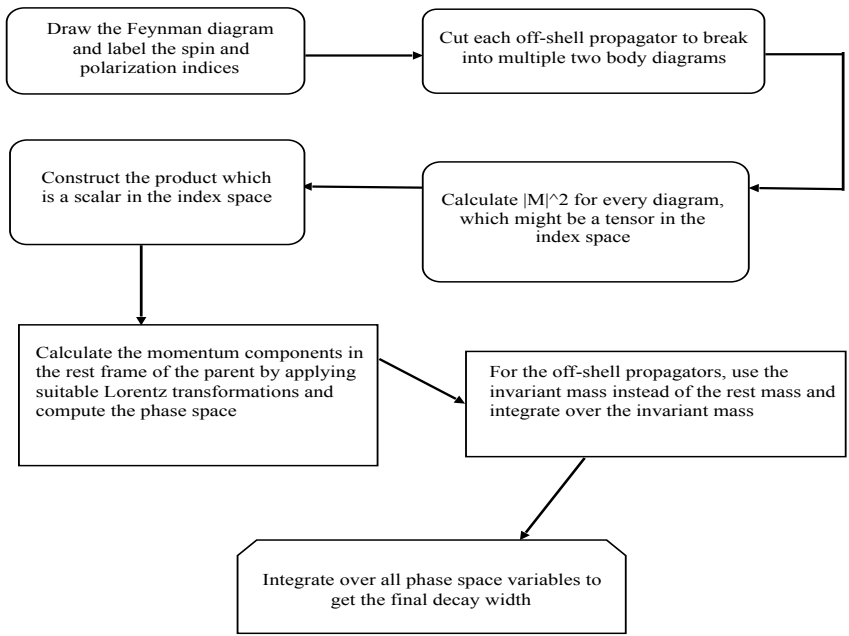}
\end{center}
\caption{\label{figflowchart} Schematic  guidelines of  the  off- shell prescription.
}
\end{figure}

For a generic $1\to n$-body decay with a single amplitude and multiple off-shell propagators, the algorithm
can be found in Ref.\ \cite{Chakrabortty:2016idh}, say in Eqs.\ (1) and (6). The essential trick is to write 
the $1\to 2$ body amplitudes in terms of the invariant masses of the off-shell legs and then integrate over 
all such invariant masses, with energy-momentum conservation. A flowchart is given in Fig.\ \ref{figflowchart}. 


Let us also mention here that for particles with relatively large decay widths, the propagator should be 
written in the Breit-Wigner form and $(m_{ij}^2-m^2)^2$ (where $m_{ij}$ is the invariant mass and $m$
is the rest mass) should be replaced by  $(m_{ij}^2-m^2)^2 + \Gamma^2 m^2$, where $\Gamma$ is the 
decay width.   
The $1\to 2$ ``decay width" $\tilde{\Gamma}$ is analogous to the actual decay width, $\Gamma$, but this is 
not a number; rather, this is a matrix in the basis of the quantum numbers carried by the off-shell 
particles. Obviously, so are the ``amplitudes" $|\mathcal{M}|^2$. 
This structure helps us to track the flow of those quantum numbers throughout the cascade.  
For a single cascade the footprints which need to be followed are structured in detail in 
Ref.\ \cite{Chakrabortty:2016idh}.

In this paper, we generalize that idea, and  also show how to take into account the full phase space,
consistent with our factorized diagrams. Often there are multiple Feynman diagrams leading to the same 
final states, and for that we have to incorporate the interference amplitudes too. This has been the 
main motivation of this paper. 
In Sections 2 and 3, we discuss the three-body and four-body decays respectively, and conclude in Section 4. 
A lot of computational details have been relegated to the Appendix.

\section{\Large{3-particle final states}}

\begin{figure}[h!]
\begin{center}
\epsfig{figure=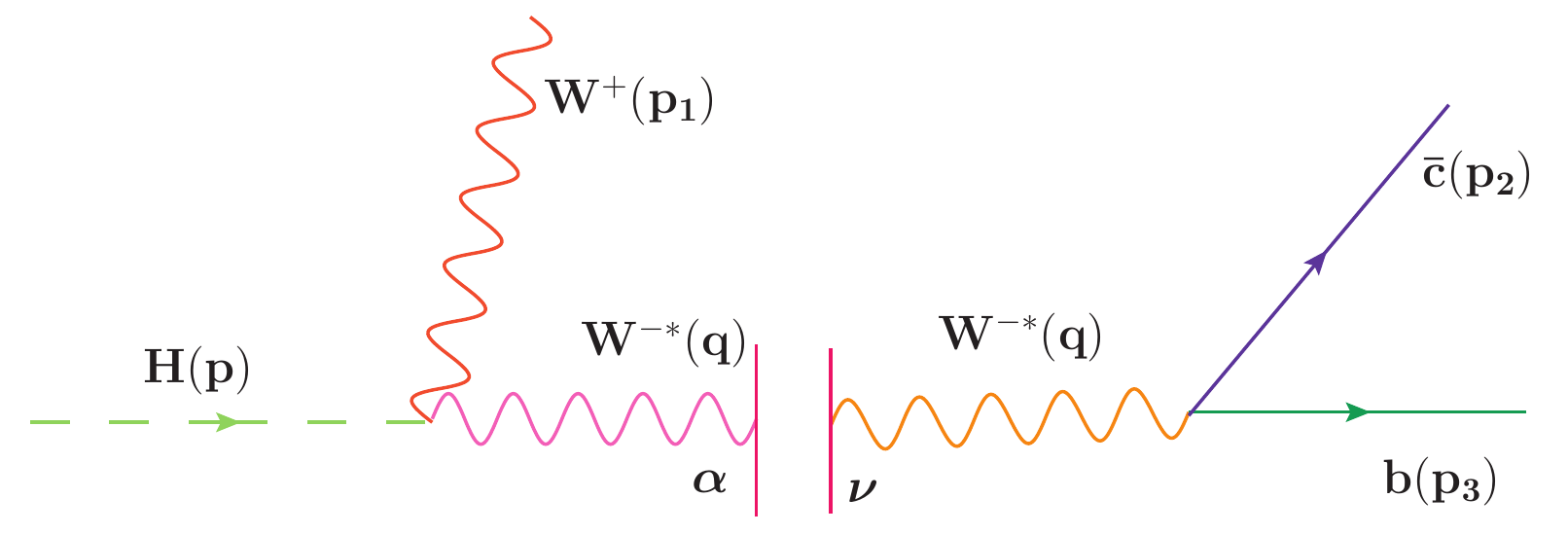, scale = 0.675}
\end{center}
\caption{\label{fig1} Cutting the off-shell propagator for the decay
$H(p) \xrightarrow{W^{+*}(q)} W^-(p_1)  c(p_2)  \bar{b}(p_3)$. }
\end{figure}

\begin{figure}[h!]
\begin{center}
\epsfig{figure=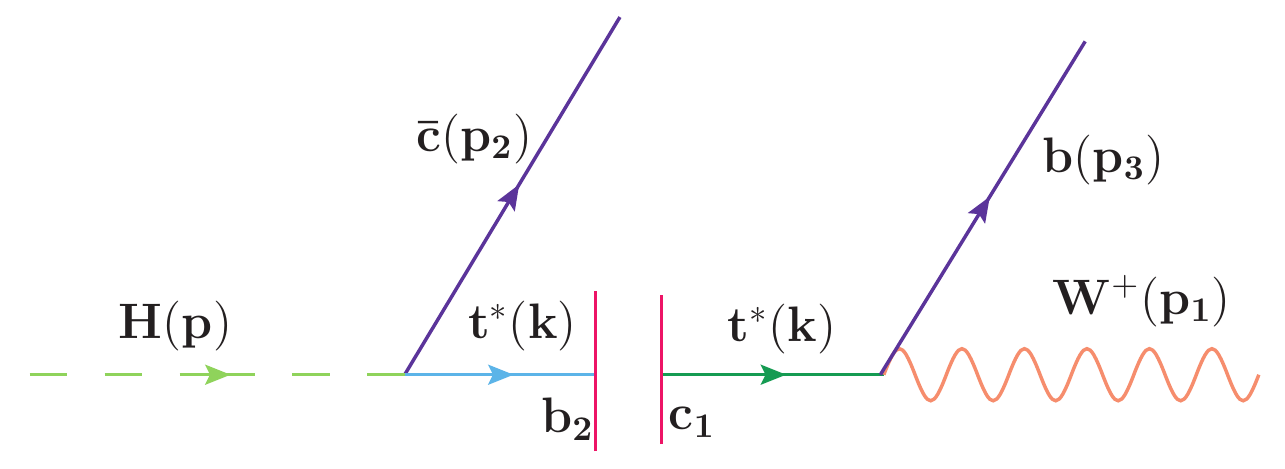, scale = 0.675}
\end{center}
\caption{\label{fig2} Cutting the off-shell propagator for the decay
$H(p) \xrightarrow{ {t}^*(k)} \bar{c}(p_2)  W^+(p_1)  \bar{b}(p_3)$. }
\end{figure}

As an example, let us first consider the 3-body decays of the neutral Higgs boson $H\to b\bar{c}W^+ $ 
through off-shell top-quark and W-boson: 
$H(p)\xrightarrow [{t}^*] {W^{-*}} W^+(p_1) \ \bar{c}(p_2) \ {b}(p_3)$. 
The relevant Feynman diagrams are shown in Figs.\ 
\ref{fig1} and \ref{fig2}. While the first amplitude is allowed in the SM itself, we have to introduce
a flavor-changing neutral current coupling for the Higgs boson to top and charm, given by 
$y_{ct}(\bar{t}c+\bar{c}t)H$, to introduce the second amplitude. The Yukawa coupling $y_{ct}$ is fixed to 
such a value as to make the amplitudes comparable in magnitude and hence emphasize the effect of the 
interference term. We take the relevant quark mixing matrix elements,
$V_{cb}$ and $V_{tb}$, and also the new Yukawa coupling $y_{ct}$, to be real. 


We will first discuss the individual contributions from these diagrams and then their interference contribution.
The amplitude squared for the decay $H(p)\xrightarrow {W^{-*}(q)}W^+(p_1) \bar{c}(p_2) {b}(p_3) $ 
has been computed in \cite{Chakrabortty:2016idh} and the result is 
\begin{equation}
\Gamma_1(H\to W^+b\bar{c}) = \frac{N_c g^2 m_W^4}{8m_H v^2} |V_{cb}|^2 
 \int\left[ \frac{1}{\pi} \frac{d m_{23}^2}{(m_{23}^2-m_W^2)^2}\right]
 \int d_{PS}^{W^*\to t b} \int d_{PS}^{H\to WW^*} {\cal F}_1\,,
\end{equation}
where $N_c=3$ is the colour factor, $g$ is the $SU(2)$ coupling, and $v$ is the vacuum expectation value 
of the Higgs field, given by $v=2m_W/g$. Furthermore, 
\begin{eqnarray} 
 {\cal F}_1 &=& \frac{16}{m_W^4}\left[ 
 2 m_{W}^2 (p_1.p_2)(p_1.p_3) + 4 m_W^2(q.p_2)(q.p_3)
 -2 m_W^2 m_{23}^2 (p_2.p_3)-2 (q.p_1)(q.p_2)(p_1.p_3) 
 \right. \nonumber\\
 && \left. 
 - 2 (q.p_1)(q.p_3)(p_1.p_2) +  2 (q.p_1)^2(p_2.p_3) +m_{23}^4 (p_2.p_3) 
 -2 m_{23}^2 (q.p_2)(q.p_3)
 \right.\nonumber\\
 && 
 \left. + 2 m_{W}^{-2}  (q.p_1)^2(q.p_2)(q.p_3) -m_{W}^{-2} m_{23}^2 (p_2.p_3)(q.p_1)^2 + m_W^4 (p_3.p_2)\right]\,. 
\end{eqnarray}
Note that we have used $q^2 = m_{23}^2$ and the factors of $m_W$ come from the polarization sum 
of the $W$-propagator.

Similarly, for the second amplitude, we get 
\cite{Chakrabortty:2016idh}.
\begin{equation}
\Gamma_2(H\rightarrow b \bar{c} W^+ ) = \Big(\frac{1}{4 m_H}\Big)\frac{N_c g^2 |y_{ct}|^2 |V_{tb}|^2}{8} \int\left[\frac{1}{\pi} \left(\frac{d m_{13}^2}{(m_{13}^2-m_t^2)^2}\right)\right]
\int d_{ps}^{H\rightarrow \bar{c} {t}^*}\int d_{ps}^{{t}^*\rightarrow {b} W} \mathcal{F}_2,
\end{equation}
where 
\begin{align}
\mathcal{F}_2=16\left((p_2.k)- m_c m_t\right)\left((k.p_3)+\frac{2(k.p_1)(p_3. p_1)}{m_W^2}\right) 
-8\left(k^2-m_t^2\right)\left( (p_2.p_3) + \frac{2(p_2.p_1)(p_3.p_1)}{m_W^2}\right).
\end{align}


Let us now concentrate on the interference diagram. 
The heart of our proposal is to decompose every cascade into several $1\to 2$ body decays 
irrespective of the length of the decay chain.
We stick to our prime intention, write down the amplitude for every two body decay, and then join them as 
shown below.
The  two body decay amplitudes for $H(p)\xrightarrow [{t}^*] {W^{-*}}  W^+(p_1) \ \bar{c}(p_2) \ {b}(p_3)$
 through off-shell $W$ and $t$, as shown in Figs.~\ref{fig1} and \ref{fig2} are written as:
\begin{align}
(\mathcal{M}_1)^{\alpha \mu}(H\rightarrow W^- W^{+*}) \ = & \ g m_W \ \epsilon^{\mu *}_{(\lambda)} (p_1) \ 
\epsilon^{\alpha *}_{(\lambda ')}(q)\,, \nonumber \\
(\mathcal{M}_2)_{\mu \nu}(W^{-*}\rightarrow \bar{c}b) \ = & \ \frac{g V_{cb}}{2\sqrt{2}}  \epsilon_{\mu(\lambda')} (q)
 \big[\bar{u}^{(s_2)}(p_3) \gamma_\nu (1-\gamma^5)  v^{(s_3)}
 (p_2)\big]\,, \nonumber \\ 
(\mathcal{M}_3)_{b_2c_1}(H\rightarrow  \bar{c} t^*) \ = & \ -y_{ct} \Big[v^{(s_3)}(p_2)\Big]_{c_1} \Big[\bar{u}^{(s)}(k)\Big]_{b_2}\,, \nonumber \\
\Big[(\mathcal{M}_4)_{c_1b_2}\Big]_\alpha^\nu(t^* \rightarrow b W^+) \ = & \ \frac{g}{2\sqrt{2}} V_{tb} \Big[ 
\bar{u}^{(s_3)}(p_3) \gamma_\alpha (1 - \gamma^5) {u}^{(s)}(k) \Big]_{c_1 b_2} \epsilon^{\nu *}_{(\lambda)}
  (p_1)\,.
\end{align}

The interference term 
$\left[\mathcal{M}_1 \mathcal{M}_2\right]^\dagger \mathcal{M}_3\mathcal{M}_4$ can be written as
\begin{align}
 &- \ \frac{N_c g^3 V_{tb}V_{cb} y_{ct} m_W}{8} \Big[\epsilon^{\mu *}_{(\lambda)} (p_1) \ {\epsilon^{\alpha *}}
 _{(\lambda ')}(q) \  \epsilon_{\mu(\lambda')} (q) \big[\bar{u}^{(s_2)}(p_3)  \gamma_\nu (1-
 \gamma^5)   v^{(s_3)}(p_2)\big]\Big]^\dagger  \nonumber \\ 
& \ \times \Big[ \big[{v}^{(s_2)}(p_2)\bar{u}^{(s)}(k) \big]_{b_2 c_1} \big[\bar{u}^{(s_3)}(p_3)  \gamma_\alpha (1 - \gamma^5) u^{(s)}(k)\big]_{c_1 b_2} \epsilon^{\nu *}_{(\lambda)} (p_1) \Big] \nonumber \\
 = & -\ \frac{N_c g^3 V_{tb}V_{cb} y_{ct} m_W}{8} \left( -g^{\mu \nu} + \frac{{p_1}^\mu p_1^\nu}{m_W^2}\right) \left( - g^{\alpha}_{\mu} +\frac{q^{\alpha} q_{\mu}}{m_W^2} \right) \nonumber\\
& \ \times \text{Tr} \Big[ (\slashed{p}_3+ m_b) \gamma_\nu (1-\gamma^5) 
 \  (\slashed{p}_2 - m_c) (\slashed{k} + m_t) \gamma_\alpha (1- \gamma^5) \Big]\nonumber \\
\ = & \ {N_c g^3 V_{tb}V_{cb} \ y_{ct} m_W} \left\{ m_t \left[\frac{2(p_1.p_3) (p_2.p_1)}{m_W^2}+
\frac{2(p_3.q) (p_2.q)}{m_W^2} + (p_3.p_2) - \frac{m_{23}^2}{m_W^2}(p_3.p_2)
\right.\right. \nonumber\\
& \ \left.\left. 
 -\frac{(p_1.q)}{m_W^4}\Big((p_1.p_3)(p_2.q)-(p_1.q)(p_3.p_2)+(p_3.q)(p_2.p_1)\Big)\right]\right.\nonumber\\
& \ \left. - m_c\Big[\frac{2(p_1.p_3) (k.p_1)}{m_W^2} + \frac{2(p_3.q) (k.q)}{m_W^2}+ (p_3.k) - \frac{m_{23}^2}{m_W^2}(p_3.k)\right.\nonumber\\
& \ \left.
 - \frac{(p_1.q)}{m_W^4}\Big((p_1.p_3)(k.q)-(p_1.q)(p_3.k)+(p_3.q)(k.p_1)\Big) \Big]\right\}.
\label{eq-int}
\end{align}

Here, we have used  the following polarization and spin sums  
\be
\sum_\lambda \epsilon_{(\lambda)}^{*\mu}(k) \epsilon_{(\lambda)}^{\nu}(k)  = -g^{\mu\nu}+\frac{k^\mu k^\nu}{m_W^2}\,,\ \ 
\sum_s u^{(s)}(p ) \bar{u}^{(s)} (p )  = (\slashed{p} + m)\,,\ \
\sum_s v^{(s)}(p ) \bar{v}^{(s)} (p )  = (\slashed{p} - m)\,.
\ee

One can easily check that this result is in complete agreement with that obtained using the full cascade 
without any decomposition. 

This is, therefore, a good place to explain the interference algorithm. The entire matrix element squared 
computed in the canonical way without cutting any off-shell propagator is obviously a Lorentz and gauge 
scalar. Here, when one calculates the individual diagrams, the $|{\cal M}|^2$ for the cut diagrams need not 
be a scalar; this can carry Lorentz or spin indices. The index contractions, as has been explained in 
Ref.\ \cite{Chakrabortty:2016idh} are performed in such a way that both the $|{\cal M}|^2$s are matrices 
in the index space but the product is a scalar. Schematically speaking, the combined $|{\cal M}|^2$ 
looks like $(|{\cal M}_1|^2)^\mu_\nu (|{\cal M}_2|^2)^\nu_\mu$, which ensures the ``index flow" through 
the cut propagator. For the interference diagrams, the Lorentz and spin indices are to be contracted in such a 
way that both $({\cal M}_1{\cal M}_2)^\dag$ and $({\cal M}_3{\cal M}_4)$ are matrices but their product 
is a scalar. The assignment can be followed in Eq.\ (\ref{eq-int}). In ${\cal M}_1{\cal M}_2$, the contraction 
of the index $\mu$ shows the ``index flow" through the off-shell $W$, and in ${\cal M}_3{\cal M}_4$, 
the spin index $b_1$ plays the same role. Note that ${\cal M}_4$ is a matrix in both spin and Lorentz spaces.
Another example of this ``index flow" is shown in the next Section.

\begin{table}[htbp]
\begin{center}
    \begin{tabular}{ | c |  c | c | c |}
    \hline
    $\Gamma_{W^{*}}$ (GeV) & $\Gamma_{t^{*}}$ (GeV) & $\Gamma_{\rm int}$ (GeV) & $\Gamma$ (GeV)
    \\ \hline
    $2.08\times 10^{-7}$(C1) & $2.49\times 10^{-7}$(C1) & $3.79\times 10^{-7}$(C1) & $8.36\times10^{-7}$ (C1)
    \\
    \hline
    $2.12\times 10^{-7}$(C2) & $2.41\times 10^{-7}$(C2) & $3.79\times 10^{-7}$(C2) & $8.32\times10^{-7}$ (C2)
   \\  \hline
    $2.11\times 10^{-7}$(M) & $2.41\times 10^{-7}$(M) & $3.79\times 10^{-7}$(M) & $8.31\times10^{-7}$ (M)
    \\
    \hline
    \end{tabular}
    \caption{\small Decay width of $H \to W^+b\bar{c}$ calculated using {\tt CalcHEP} v3.6.27 (C1), 
    v3.6.23 (C2), and our algorithm with phase space integration numerically performed by 
    {\tt Mathematica} v10 (M).  }
\end{center}
\end{table}

After implementing the  three body phase space, as discussed in 
Section \ref{sec1},  we can add up all three contributions and write down the  full partial decay width as 
$\Gamma=\Gamma_1+\Gamma_2+\Gamma_{\text{int}}$, where 
the interference contribution is
\begin{eqnarray}
\Gamma_\text{int}\left( H\to W^+ b \bar{c}\right) &=& 
\frac{1}{ m_H} \left[\int \frac{1}{\pi} \left(\frac{d m_{23}^2}{(m_{23}^2-m_W^2)(m_{13}^2-m_t^2)}\right)\right]
\times \left[ \frac{1}{2}\int \frac{\bar{\beta}}{8\pi}\frac{d \cos\theta}{2}\frac{d\phi}{2\pi}\right] \times
\nonumber\\
&& \left[\frac{1}{2}\int \frac{\bar{\beta}_{23}}{8\pi}\frac{d \cos\theta_{23}}{2}\frac{d\phi_{23}}{2\pi} \right]
\times \left[ 2 \text{Re}\left( [\mathcal{M}_1 \mathcal{M}_2]^\dagger \mathcal{M}_3\mathcal{M}_4\right) 
\right]\,,
\end{eqnarray} 
where the notations have been explained in the Appendix. 

We have computed the total decay width $\Gamma$ for this process,  
using {\tt Mathematica} v10 \cite{Mathematica} to integrate the phase space numerically. 
The couplings we use are given by\footnote{Those not shown here are fixed to their SM values.}
$V_{tb} = 1$, $V_{cb}=0.04$, $y_{ct}=0.04$, 
and the masses (in GeV) are $m_H = 125$ GeV, $m_c = 1.2$ GeV, 
$m_b = 4.23$ GeV, $m_W=80.385$ GeV, and $m_t = 172.5$ GeV. 
The result has been compared with that obtained from {\tt CalcHEP} v3.6.23 as well as 
v3.6.27 \cite{Belyaev:2012qa}, 
which does the phase space integration with a numerical simulation. The comparison is shown in 
Table 1; 
we find an excellent agreement within the error margin.

\section{4-particle final states}

\begin{figure}[h!]
\begin{center}
\epsfig{figure=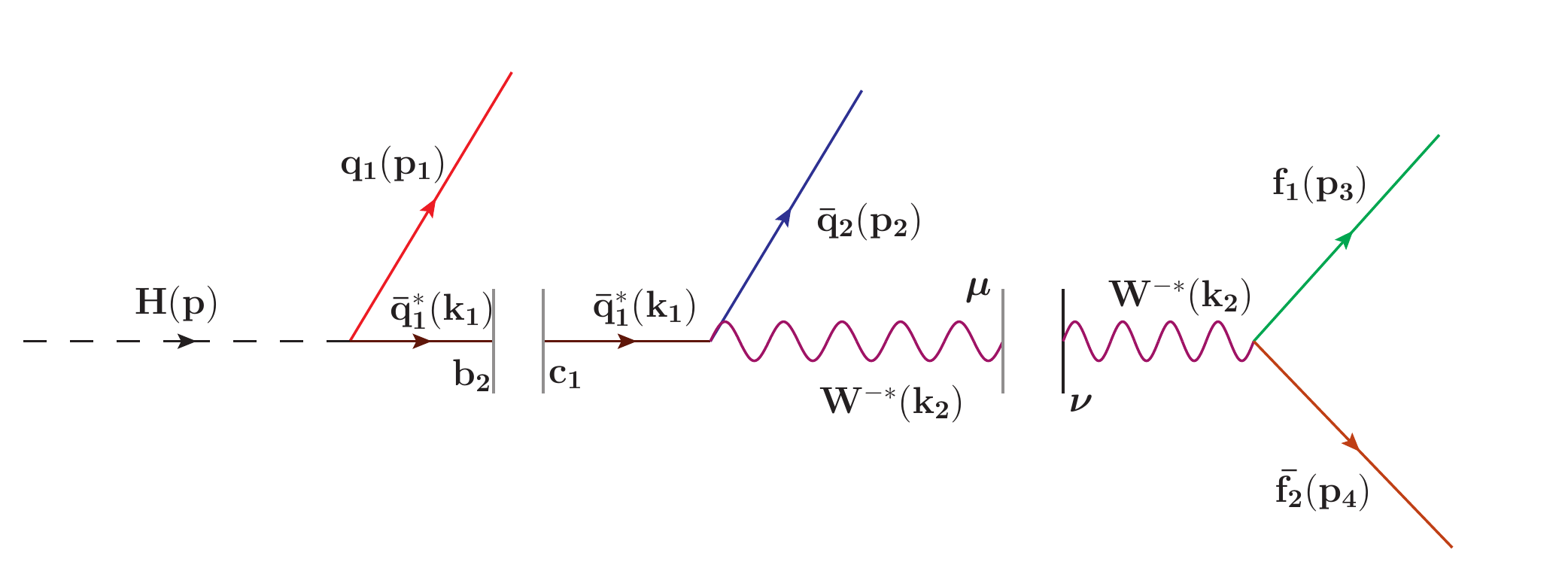,height=5cm}
\end{center}
\caption{\label{fig04} Cutting of two off-shell propagators for the decay
$H\xrightarrow{\bar{q}_1,W} q_1 \bar{q}_2 f_1 \bar{f}_2 $.
}
\end{figure}

\begin{figure}[h!]
\begin{center}
\epsfig{figure=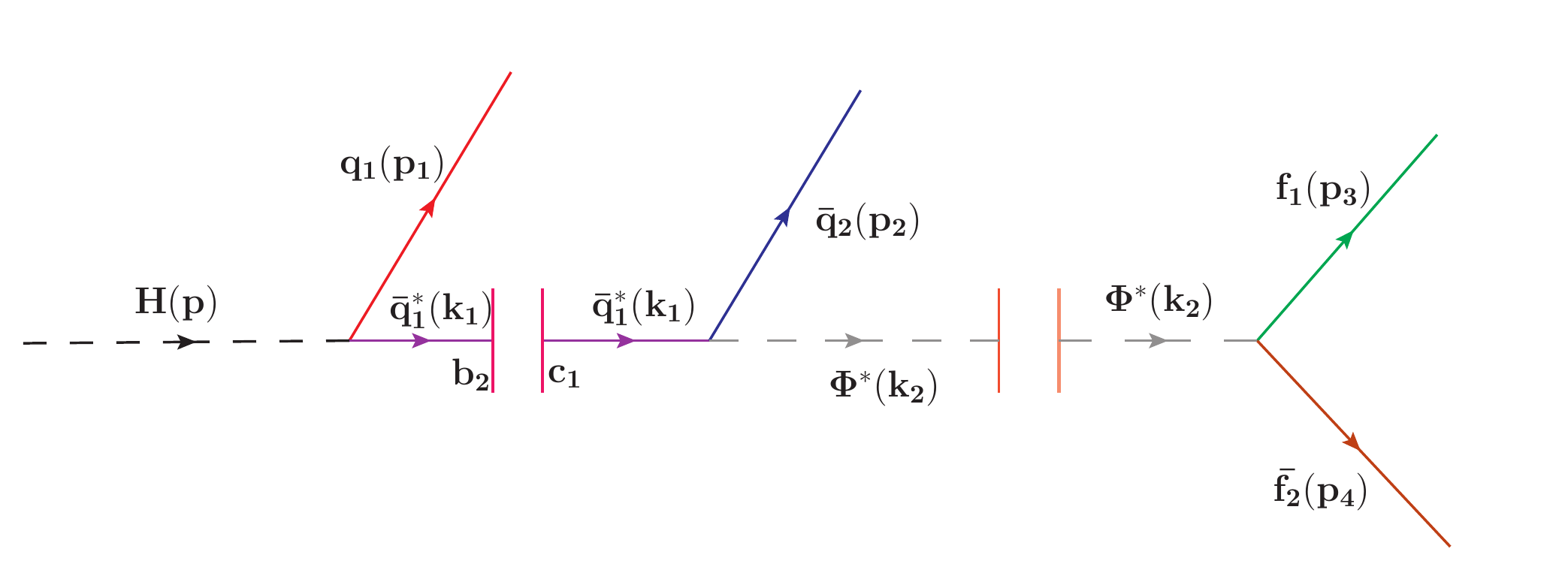,height=5.00cm}
\end{center}
\caption{\label{fig05}  Cutting of two off-shell propagators for the decay
$H\xrightarrow{\bar{q}_1,\Phi} q_1 \bar{q}_2 f_1 \bar{f}_2 $.}
\end{figure}

As an example of a $1\to 4$ decay, let us consider the decay
$H( p )  \to q_1(p_1)\, \bar{q}_2(p_2)\, f_1(p_3)\, \bar{f}_2(p_4)$ where $q_1$, $q_2$, $f_1$ and $f_2$ 
are four fermions, possibly quarks. The first amplitude proceeds through $H\to q_1 \bar{q}_1^*$, $\bar{q}_1^*
\to \bar{q}_2 W^*$, $W^*\to f_1 \bar{f}_2$. To get the second amplitude, we introduce a hypothetical 
charged scalar $\Phi$ that replaces the $W$ boson in the first amplitude. The corresponding Feynman 
diagrams are shown  in Figs.\ \ref{fig04} and \ref{fig05}.


According to our proposal the decay width for the diagram in Fig.~\ref{fig04} can be written as:
\bea
\label{e1}
\Gamma_1(H\rightarrow q_1 \bar{q}_2 f_1 \bar{f}_2) & = &
\frac{1}{m_H}\int \left[ \frac{1}{\pi} \left(  \frac{d m_{12}^2}{({m^2_{12}} - {m^2_{q_1}})^2}\right) 
\right] \left[ \frac{1}{\pi} \left(  \frac{d m_{23}^2}{({m^2_{23}} - {m^2_W})^2}\right) \right]\times \nonumber \\
&& \; \text{Tr}\left[\tilde{\Gamma}_1(H\rightarrow q_1 \bar{q}_1^*)\tilde{\Gamma}_2 (\bar{q}_1^*\rightarrow 
\bar{q}_2 W^*)\tilde{\Gamma}_3 (W^* \rightarrow f_1 \bar{f}_2)\right]\,,
\eea
with
\begin{align}
\label{e2}
&\left[\tilde{\Gamma}_1 (H\rightarrow q_1 \bar{q}_1^*)\right]_{c_1 b_2}\ 
=\ \int \frac{d_{PS}^{H\rightarrow q_1 \bar{q}_1^*}}{2}
\left[|\mathcal{M}_1(H\rightarrow q_1 \bar{q}_1^*)|^2\right]_{c_1 b_2}\,, \nonumber \\
&\left[ \left[\tilde{\Gamma}_2(\bar{q}_1^*\rightarrow \bar{q}_2 W^*)\right]_{b_2 c_1}\right]_\nu^\mu \ 
=\ \int \frac{d_{PS}^{\bar{q}_1^*\rightarrow \bar{q}_2 W^*}}{2}\left[ \left[|\mathcal{M}_2
(\bar{q}_1^*\rightarrow \bar{q}_2 W^*)|^2\right]_{b_2 c_1}\right]_\nu^\mu \,, \nonumber\\
&\left[ \tilde{\Gamma}_3 (W^* \rightarrow f_1 \bar{f}_2)\right]_\mu^\nu \ =\ 
\int \frac{d_{PS}^{W^* \rightarrow f_1 \bar{f}_2}}{2}\left[|\mathcal{M}_3(W^* \rightarrow f_1 \bar{f}_2)|^2
\right]_\mu^\nu\,,
\end{align}
where the squared amplitudes are
\begin{align}
\left[\vert\mathcal{M}_1\vert^2\right]_{c_1 b_2} \ = & \ 
\frac{N_c g^2 m_{q_1}^2}{4 m_W^2} \left\{ \left[\bar{u}^{(s_1)}(p_1)\right]_{c_1}
\left[ v^{(s_2)}(k_1)\right]_{b_2}\left[ \bar{v}^{(s_2)}(k_1)\right]_{d_1}\left[u^{(s_1)}(p_1)\right]_{d_1}\right\}\,,
\nonumber\\
\left[ \left[\vert \mathcal{M}_2 \vert^2\right]_{b_2 c_1}\right]_\nu^\mu \ = & 
\ \frac{g^2 |V_{q_1q_2}|^2}{8}\left\{  \left[ \bar{v}^{(s_3)}(k_1)\right]_{a_1} \left[ \gamma^\mu 
(1-\gamma^5)\right]_{a_1 a_2} \left[ v^{(s_4)}(p_2)\right]_{a_2} \times \right. \nonumber \\ 
\ & \left. \left[ \bar{v}^{(s_4)}(p_2)\right]_{b_1} \left[ \gamma^\alpha (1-\gamma^5)\right]_{b_1 b_2} 
\left[v^{(s_3)}(k_1)\right]_{c_1} \right\} \epsilon_\alpha^{(\lambda)} (k_2) \epsilon_\nu^{(\lambda) *} (k_2)\,, 
\nonumber\\
\left[ \vert \mathcal{M}_3\vert^2 \right]_\mu^\nu \ = 
& \ \frac{N_c g^2 |V_{f_1 f_2}|^2}{8}\left\{ \bar{u}^{(s_5)}({p}_3)\gamma_\mu (1-\gamma^5) 
v^{(s_6)}({p}_4) \bar{v}^{(s_6)}({p}_4)\gamma_\beta (1-\gamma^5) u^{(s_5)}({p}_3)\right\}\times \nonumber\\  
&   \;  \epsilon^{\beta *}_{(\lambda)} (k_2) \epsilon^\nu_{(\lambda)} (k_2)\,.
\label{e3}
\end{align}

We can now combine, at a step, two such squared amplitudes:
\bea
\left[\vert \mathcal{M}_{12}\vert^{2}\right]_\nu^\mu &= &
\left[\vert\mathcal{M}_1\vert^2\right]_{c_1 b_2} \left[ \left[\vert\mathcal{M}_2\vert^2\right]
_{b_2 c_1}\right]_\nu^\mu \nonumber\\
& = &  \frac{N_c g^4 m_{q_1}^2|V_{q_1 q_2}|^2}{32 m_W^2} g_{\alpha\nu}\times\nonumber\\
&&
\text{Tr} \left[ (\slashed{k_1} - m_{q_1})(\slashed{p_1} + m_{q_1})(\slashed{k_1} - m_{q_1})
\gamma^\mu (1 - \gamma^5) (\slashed{p_2}-m_{q_2}) \gamma^\alpha(1-\gamma^5) \right] \nonumber\\ 
& = &\frac{N_c g^4 m_{q_1}^2 |V_{q_1 q_2}|^2}{32 m_W^2} g_{\alpha \nu} \times\nonumber\\
&& \left[ 16\left( (k_1 . p_1) - m_{q_1}^2 \right) \left[ k_1^\mu p_2^\alpha - g^{\mu \alpha} (k_1 . p_2) 
+ k_1^\alpha p_2^\mu  \ + i (k_1)_\rho (p_2)_\delta \; \epsilon^{\rho \mu \delta \alpha} \right]\right. \times 
\nonumber \\ 
&& \left. +8\left( m_{q_1}^2 - {k_1}^2 \right) \left[ p_1^\mu p_2^\alpha - g^{\mu \alpha} (p_1 . p_2) + 
p_1^\alpha p_2^\mu  \ + i (p_1)_\rho (p_2)_\delta \; \epsilon^{\rho \mu \delta \alpha} \right]  \right]\,,
\eea
and with 
\be
\left[\vert\mathcal{M}_3\vert^2 \right]_\mu^\nu =  N_c g^2 |V_{f_1 f_2}|^2 g^{\nu \beta} \left[ 
{(p_3)}_\beta {(p_4)}_\mu - g_{\beta \mu} (p_3 . p_4) +  {(p_3)}_\mu {(p_4)}_\beta  \ + 
i {(p_3)}^\tau {(p_4)}^\lambda \epsilon_{\tau \beta \lambda \mu} \right]\,,
\ee
one gets $|\mathcal{M}|^2$ as:
\begin{align}
\label{e4}
 \text{Tr} \left[\vert\mathcal{M}_{12}\vert^{2} \vert\mathcal{M}_3\vert^2 \right] = 
 \frac{N_c^2 g^6 m_{q_1}^2 |V_{q_1 q_2}|^2 |V_{f_1 f_2}|^2}{ m_W^2} \left[ 
 2(k_1 . p_1 -  m_{q_1}^2){(k_1. p_3)(p_2 . p_4)} +(m_{q_1}^2-k_1^2)(p_1. p_3)(p_2 . p_4)\right]\,.
\end{align}

This expression completely agrees with the canonically computed expression, say as in Ref.\ 
\cite{Grau:1990uu}.
The phase space is incorporated as shown in Eqs.\ 
 (\ref{e_2.2.1}), (\ref{e_2.2.2}), (\ref{e_2.2.3}), (\ref{e_2.2.4}), (\ref{e_2.2.5}) and (\ref{e_2.2.6}).

The decay width is now trivial to write down:
\bea
\Gamma_1(H\rightarrow q_1 \bar{q}_2 f_1 \bar{f}_2) &=&
 \frac{1}{m_H}\int \left[ \frac{1}{\pi} \left(  \frac{d m_{12}^2}{({m^2_{12}} - {m^2_{q_1}})^2}\right) \right]
 \left[ \frac{1}{\pi} \left(  \frac{d m_{23}^2}{({m^2_{23}} - {m^2_W})^2}\right) \right]\times \nonumber \\
&& \left( \frac{1}{2}\int \frac{\bar{\beta}_1}{8\pi}\frac{d \cos\theta}{2}\frac{d\phi}{2\pi}\right)
\left( \frac{1}{2}\int \frac{\bar{\beta}_{2}}{8\pi}\frac{d \cos\theta_{1}}{2}\frac{d\phi_{1}}{2\pi}\right)
\left( \frac{1}{2}\int \frac{\bar{\beta}_{3}}{8\pi}\frac{d \cos\theta_{2}}{2}\frac{d\phi_{2}}{2\pi}\right)\times 
\nonumber \\
&& 
 \rm{Tr} \left[\left\vert\mathcal{M}_{12}\right\vert^{2} \left\vert\mathcal{M}_3\right\vert^2 \right]\,.
 \eea

The second diagram, as shown in Fig.\ \ref{fig05}, is analogous with $W$ replaced by a hypothetical charged 
scalar $\Phi$, whose coupling to any fermionic pair $pq$ is written as $y_{pq}$. As before, 
\bea
\Gamma_2(H\rightarrow q_1 \bar{q}_2 f_1 \bar{f}_2) & = &
\frac{1}{m_H}\int \left[ \frac{1}{\pi} \left(  \frac{d m_{12}^2}{({m^2_{12}} - {m^2_{q_1}})^2}\right) \right]
\left[ \frac{1}{\pi} \left(  \frac{d m_{23}^2}{({m^2_{23}} - {m^2_\Phi})^2}\right) \right] \times \nonumber \\
&& \text{Tr} \left[\tilde{\Gamma}_1 (H\rightarrow q_1 \bar{q}_1^*)\, 
\tilde{\Gamma}_2 (\bar{q}_1^*\rightarrow \bar{q}_2 \Phi^*) \, 
\tilde{\Gamma}_3 (\Phi^* \rightarrow f_1 \bar{f}_2)\right]\,,
\eea
with
\begin{align}
&\left[\tilde{\Gamma}_1(H\rightarrow q_1 \bar{q}_1^*)\right]_{c_1 b_2}\ =
\ \int \frac{d_{PS}^{H\rightarrow q_1 \bar{q}_2^*}}{2}
\left[|\mathcal{M}_1(H\rightarrow q_1 \bar{q}_2^*)|^2\right]_{c_1 b_2}\,, \nonumber \\
& \left[\tilde{\Gamma}_2 (\bar{q}_1^*\rightarrow \bar{q}_2 \Phi^*)\right]_{b_2 c_1} \ =
\ \int \frac{d_{PS}^{\bar{q}_1^*\rightarrow \bar{q}_2 \Phi^*}}{2} 
\left[|\mathcal{M}_2(\bar{q}_1^*\rightarrow \bar{q}_2 \Phi^*)|^2\right]_{b_2 c_1}\,,\nonumber\\
& \tilde{\Gamma}_3 (\Phi^* \rightarrow f_1 \bar{f}_2) \ =
\ \int \frac{d_{PS}^{\Phi^* \rightarrow q_2 \bar{q}_3}}{2} |\mathcal{M}_3(\Phi^* \rightarrow f_1 \bar{f}_2)|^2\,.
\end{align}

These functions contain the squared amplitudes for the $1\to 2$-body processes:
\begin{align}
\left[ |\mathcal{M}_1|^2 \right]_{c_1 b_2} \ = 
& \ \frac{N_c g^2 m_{q_1}^2}{4 m_W^2}
\left\{ \left[u^{(s_1)}(p_1)\right]_{c_1}\left[ v^{(s_2)}(k_1)\right]_{d_1}\left[ \bar{v}^{(s_2)}(k_1)\right]_{b_2}\left[\bar{u}^{(s_1)}(p_1)\right]_{d_1}
\right\}\,,\nonumber\\
\left[ |\mathcal{M}_2|^2 \right]_{b_2 c_1} \ = 
& \ |y_{q_1q_2}|^2 \left\{ \left[v^{(s_1)}(p_2)\right]_{b_2}\left[ v^{(s_2)}(k_1)\right]_{d_1}\left[ \bar{v}^{(s_2)}(k_1)\right]_{c_1}\left[\bar{v}^{(s_1)}(p_2)\right]_{d_1}
\right\}\,, \nonumber\\
\left[ |\mathcal{M}_3|^2 \right] \ = & \ 4N_c |y_{f_1 {f}_2}|^2 \left\{ \left[u^{(s_1)}(p_3)\right]_{a_2}\left[ v^{(s_2)}(p_4)\right]_{a_1}\left[ \bar{v}^{(s_2)}(p_4)\right]_{a_2}\left[\bar{u}^{(s_1)}(p_3)\right]_{a_1}
\right\}\, \,.
\end{align}

As the scalar propagator does not carry any polarization index, $|{\cal M}_3|^2$ is only a number. 
Combining all the squared amplitudes, we get
\bea
\label{e5}
\left|\mathcal{M}_1|^2 \right]_{c_1 b_2} \left[ |\mathcal{M}_2|^2 \right]_{b_2 c_1} 
|\mathcal{M}_3|^2 
 & = & \frac{N_c^2 g^2 m_{q_1}^2 |y_{q_1q_2}|^2 |y_{f_1 {f}_2}|^2}{m_W^2}\,
 \left[ (p_3 . p_4) - m_{f_1} m_{f_2} \right]\times\nonumber\\
 &&  \left[ 8(k_1 . p_1)(k_1 . p_2)  -4k_1^2(p_2.p_1) - 8 m_{q_1}^2(p_2.k_1) + 4 m_{q_1}^2(p_2.p_1)
 \right.\nonumber\\
 &&\left. -4 m_{q_2}\left(m_{q_1}k_1^2 -2m_{q_1}(p_1.k_1)+m_{q_1}^3 \right) \right]\,. 
\eea

 The decay width, therefore, is 
 \bea
 \Gamma_2(H\rightarrow q_1 \bar{q}_2 f_1 \bar{f}_2) &=&
 \frac{1}{m_H}\int \left[ \frac{1}{\pi} \left(  \frac{d m_{12}^2}{({m^2_{12}} - {m^2_{q_1}})^2}\right) \right]
 \, \left[ \frac{1}{\pi} \left(  \frac{d m_{23}^2}{({m^2_{23}} - {m^2_\Phi})^2}\right) \right]\times\nonumber \\
&& \left( \frac{1}{2}\int \frac{\bar{\beta}_1}{8\pi}\frac{d \cos\theta}{2}\frac{d\phi}{2\pi}\right)\,
\left( \frac{1}{2}\int \frac{\bar{\beta}_{2}}{8\pi}\frac{d \cos\theta_{1}}{2}\frac{d\phi_{1}}{2\pi}\right)\,
\left( \frac{1}{2}\int \frac{\bar{\beta}_{3}}{8\pi}\frac{d \cos\theta_{2}}{2}\frac{d\phi_{2}}{2\pi}\right)\times \nonumber \\
&&  \text{Tr} \left[ |\mathcal{M}_1|^2\,  |\mathcal{M}_2|^2\,  |\mathcal{M}_3|^2 \right]\,.
\eea

The interference term has to be calculated following the ``index flow" algorithm as discussed for the $1\to 3$ 
decays. The spin indices are assigned as shown in Figs.\ \ref{fig04} and \ref{fig05}. The $1\to 2$ 
amplitudes are as follows:

\begin{align}
(\mathcal{M}_1)_{b_2a_1}(H\rightarrow q_1 \bar{q}_1^*) \ = & \ 
-\frac{g}{2}\frac{m_{q_1}}{m_W} \big[ \bar{u}^{(s_1)}(p_1)\big]_{b_2}  \Big[ v^{(s)} (k_1)\big]_{ a_1}\,, \nonumber\\
\Big[(\mathcal{M}_2)_{a_1b_2}\Big]^\mu_\nu(\bar{q}_1^* \rightarrow \bar{q}_2 W^*) \ = & \ \frac{g}{2\sqrt{2}} V_{q_1 q_2} \big[ \bar{v}^{(s)}(k_1)  \gamma^\mu (1-\gamma^5)  v^{(s_2)}(p_2)\big]_{a_1 b_2} \epsilon^{(\lambda)*}_\nu (k_2)\,, \nonumber\\
(\mathcal{M}_3)^\nu_\mu (W^* \rightarrow f_1 \bar{f}_2) \ = & \ \frac{g V_{f_1 f_2}}{2\sqrt{2}} \epsilon_\mu^{(\lambda)}(k_2)\big[ \bar{u}^{(s_3)}(p_3) \gamma^\nu (1-\gamma^5)   v^{(s_4)}(p_4)\big]\,, \nonumber\\
(\mathcal{M}_4)_{c_1d_1}(H\rightarrow q_1 \bar{q}_1^*) \ = & -\ \frac{g}{2}\frac{m_{q_1}}{m_W} \big[ \bar{u}^{(s_1)}(p_1)\big]_{c_1} \big[ v^{(s)} (k_1)\big]_{ d_1}\,, \nonumber\\
(\mathcal{M}_5)_{d_1c_1}(\bar{q}_1^* \rightarrow \bar{q}_2 \Phi^*) \ = & \ -y_{q_1 q_2} \big[ \bar{v}^{(s)}(k_1)\big]_{d_1} \big[v^{(s_2)}(p_2)\big]_{c_1}\,, \nonumber\\
\mathcal{M}_6 (\Phi^* \rightarrow f_1 \bar{f}_2 ) \ = & \ -y_{f_1 {f}_2} \big[ \bar{u}^{(s_3)}(p_3) v^{(s_4)}(p_4)\big]\,.
\end{align}

Combining all these contributions, the interference term comes out to be 
\begin{align}
& \rm{Tr}\left[\mathcal{M}_3^\dag \mathcal{M}_2^\dag \mathcal{M}_1^\dag 
\mathcal{M}_4 \mathcal{M}_5 \mathcal{M}_6\right] \nonumber \\
\ = & \ \frac{N_c^2 g^4}{32} \frac{m_{q_1}^2}{m_W^2} V_{q_1 q_2} y_{q_1 q_2} y_{f_1 {f}_2}
\left( \bar{u}^{(s_3)}(p_3)v^{(s_4)}(p_4)\right)\times \nonumber\\
 & \left(\epsilon_\mu^{(\lambda)}(k_2) \bar{u}^{(s_3)}(p_3)\gamma^\nu (1-\gamma^5)   v^{(s_4)}(p_4) \right)^\dag \times
 \nonumber \\
&  \left( \left[ \bar{v}^{(s)}(k_1)  \gamma^\mu (1-\gamma^5)  v^{(s_2)}(p_2)\right]_{a_1b_2} \epsilon^{(\lambda)*}_\nu (k_2)\right)^\dag \times  \nonumber \\
& \left( \left[ \bar{u}^{(s_1)}(p_1) \right]_{b_2} \left[v^{(s)} (k_1)\right]_{a_1} \right)^\dag
 \left(\left[ \bar{u}^{(s_1)}(p_1)\right]_{c_1}  \left[v^{(s)} (k_1)\right]_{ d_1} \right)  
 \left( \left[ \bar{v}^{(s)}(k_1)\right]_{d_1} \left[v^{(s_2)}(p_2)\right]_{c_1}\right)\nonumber \\
\ = & - \frac{N_c^2 g^4}{32} \frac{m_{q_1}^2}{m_W^2} V_{q_1 q_2} y_{q_1q_2} y_{f_1 f_2} 
{\rm Tr}\left[ ({\slashed p}_3 + m_{f_1})(\slashed{p}_4 - m_{f_2})\gamma^\mu(1-\gamma^5)\right]
\times \nonumber \\
& {\rm Tr} \left[ (\slashed{p}_2 - m_{q_2})\gamma_\mu (1-\gamma^5)(\slashed{k}_1 - m_{q_1})
(\slashed{p}_1 + m_{q_1}) (\slashed{k}_1 - m_{q_1}) \right]\nonumber\\
=& \frac{N_c^2 g^4m_{q_1}^2}{2m_W^2} V_{q_1 q_2} y_{q_1q_2} y_{f_1f_2} 
\left[ (m_{q_1}^3+m_{q_1} k_1^2 - 2 m_{q_1}(p_1.k_1)) \left\{ m_{f_2}(p_2.p_3)-m_{f_1}(p_2.p_4) \right\} -
m_{q_2}(m_{q_1}^2-k_1^2) \right. \times\nonumber \\ 
& \left. \left\{ m_{f_2}(p_1.p_3)-m_{f_1}(p_1.p_4) \right\} +2m_{q_2}(m_{q_1}^2-p_1.k_1)
\left\{ m_{f_2}(k_1.p_3)-m_{f_1}(k_1.p_4) \right\}\right]\,.
\end{align}

After  incorporating  the four-body phase space structure as shown in Section~\ref{Asym_Decay}, 
the contribution to the decay width from the interference diagram is
\begin{align}
\Gamma_\text{int}(H\rightarrow q_1 \bar{q_2} f_1 \bar{f}_2) = & 
\frac{1}{m_H}\int \left[ \frac{1}{\pi} \left(  \frac{d m_{12}^2}{({m^2_{12}} - {m^2_{q_1}})^2}\right) \right]
\left[ \frac{1}{\pi} \left(  \frac{d m_{23}^2}{({m^2_{23}} - {m^2_W}) ({m^2_{23}} - {m^2_\Phi})}\right) \right]
\times\nonumber \\
& \left(\frac{1}{2}\int \frac{\bar{\beta}_1}{8\pi}\frac{d \cos\theta}{2}\frac{d\phi}{2\pi}\right)
\left( \frac{1}{2}\int \frac{\bar{\beta}_{2}}{8\pi}\frac{d \cos\theta_{1}}{2}\frac{d\phi_{1}}{2\pi}\right)
\left( \frac{1}{2}\int \frac{\bar{\beta}_{3}}{8\pi}\frac{d \cos\theta_{2}}{2}\frac{d\phi_{2}}{2\pi}\right) 
\times \nonumber \\
& 
\left[  2~{\rm Re}(\mathcal{M}_6\mathcal{M}_3^\dagger \mathcal{M}_2^\dagger \mathcal{M}_1^\dagger \mathcal{M}_4 \mathcal{M}_5)\right]\,.
\end{align}

The total decay width for the process is given by
\be
\Gamma=\Gamma_1+\Gamma_2+\Gamma_\text{int}\,.
\ee

As an example, we consider the decay $H\to t\bar{b} b\bar{c}$, keeping the top mass intentionally light enough for 
this decay to be kinematically possible but making sure that $2m_t > m_H$. 
The comparison with {\tt CalcHEP}, as shown in Table \ref{tab-4body}, is quite impressive. For evaluation, 
we have used 
  $V_{tb}=1$, $V_{cb}=0.04$, $y_{tb}=y_{cb}=1$, $m_h=125$ GeV, $m_t=65$ GeV, and $m_W=80.385$ GeV, 
  varying the mass of $\Phi$. Note that we could have had
  two amplitudes even without the introduction of $\Phi$ through a symmetric cascade: $H\to W^{+*}W^{-*}$, 
  $W^{+*}\to t\bar{b}$, $W^{-*}\to b\bar{c}$, by suitable adjusting the masses. Such a symmetric cascade
  has already been discussed in Ref.\ \cite{Chakrabortty:2016idh}.
    
  \begin{table}[htbp]
\begin{center}
    \begin{tabular}{ | c | c | c | c |}
    \hline
 $m_t$ & $m_\Phi$ & $\Gamma~({\rm C})$ & $\Gamma~{\rm (M)}$\\
 \hline
  65 & 200 & $1.33\times 10^{-8}$ & $1.47\times 10^{-8}$ \\
      65 & 150 & $4.36\times 10^{-8}$ & $4.87\times 10^{-8}$ \\
    65 & 100 & $2.51\times 10^{-7}$ & $2.79\times 10^{-7}$ \\
  \hline
    \end{tabular}
     \caption{Comparison of the decay width for the process $H\to t\bar{b}b\bar{c}$ with 
     {\tt CalcHEP} v3.6.23 (denoted by C) and {\tt Mathematica} v10 (denoted by M)
     using our algorithm. All entries are in GeV. }
    \label{tab-4body}
\end{center}
\end{table}

\section{Summary and conclusion}

In this paper we extend and generalize our algorithm to extend its applicability to processes with 
multiple amplitudes and hence with interference contributions in squared amplitudes. One can decompose 
the entire chain in several $1\to 2$ decays with off-shell particles as incoming and/or outgoing legs. 
The algorithm for ``index flow" for the interference diagrams has been exemplified by a $1\to 3$ and a 
$1\to 4$ decay process. The detailed calculation of the phase space has been discussed in the Appendix.

We find an impressive agreement with the results obtained with the software {\tt CalcHEP} that calculates 
the phase space by Monte Carlo simulation. This shows that one can reach about the same level of accuracy 
with much less computer time. The present paper completes our discussion on tree-level decays. 

Another advantage of the algorithm is the way one can reduce it to a limited number of elementary 
vertices. There are only six such types: $\phi\phi\phi$, $VVV$, $\phi\phi V$, $\phi VV$,
$f\bar{f}\phi$ and $f\bar{f}V$, where $\phi$, $V$, and $f$ stand for any generic spin-0, spin-1, and 
spin-$\frac12$ particle. Once one specifies the particle content of each vertex and the momentum flow, 
the entire cascade can be built up using those $1\to 2$ diagrams as building blocks. Following this 
technique, we plan to automatize the algorithm in near future.

\section*{Acknowledgement}
The authors would like to acknowledge Palash B.\ Pal for some interesting discussions which led to the 
present work. 
J.C.\ is supported by the Department of Science and Technology, Government of India, 
under the Grant IFA12-PH-34 (INSPIRE Faculty Award); and the Science and Engineering 
Research Board, Government of India, under the agreement SERB/PHY/2016348.  
A.K.\ acknowledges the Council for Scientific and
Industrial Research, Government of India, for a research grant. 
R.M.\ acknowledges the University Grants Commission, Government of India, for providing financial support 
through UGC-CSIR NET JRF.


\appendix
\numberwithin{equation}{section}

\section*{APPENDIX}

\section{Phase space decomposition}\label{phase_space}

Our proposal is based on the decomposition of a long cascade to several $1\to 2$ subprocesses and then 
putting them together following the algorithm proposed. Here, we show how the full phase space may
look like when we decompose that in terms of two-body phase spaces. 
Interference diagrams are also included in the discussion. 
We adopt some toy decays without specifying the quantum numbers for the off-shell propagators.

\subsection{\large{\textbf{Three-body decay}}}\label{sec1}
\begin{figure}[h!]
\begin{center}
\epsfig{figure=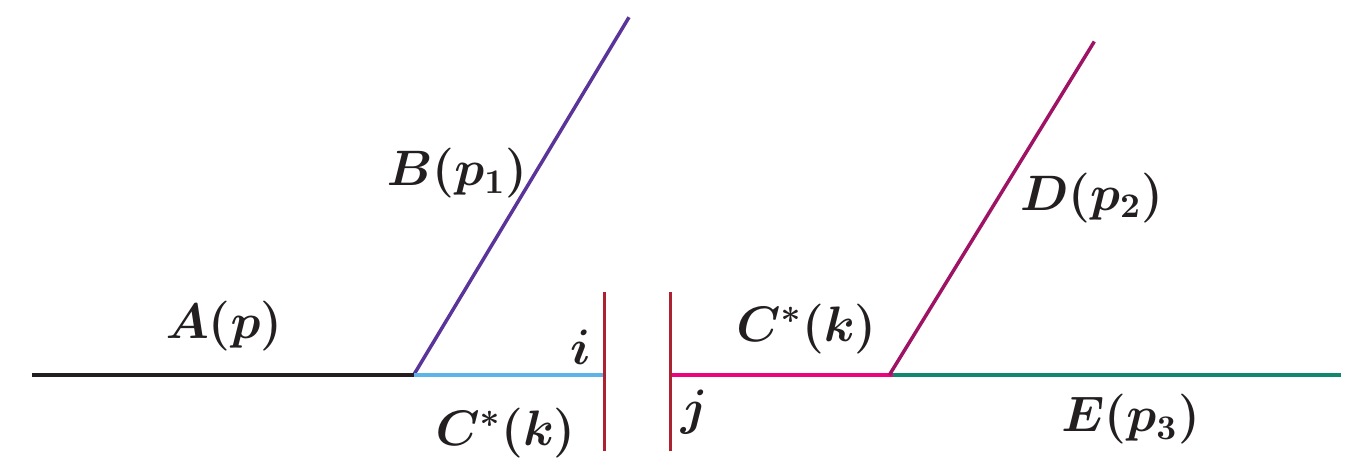,scale=0.575}
\end{center}
\caption{\label{fig1.1}  Representative figure of the three-body decay, $A \rightarrow B D E  $.
}
\end{figure}
Let us consider a toy three-body decay: $A( p )\to B(p_1) C^*(k)\to B(p_1) D(p_2) E(p_3)$, and also assume 
that  $A$  is a scalar while particles $B$ to $E$ can have any spin. 
According to our proposal the decay width for this process can be written \cite{Chakrabortty:2016idh} as
\begin{eqnarray}
\Gamma(A\rightarrow B D E)=\frac{f_s }{m_A}\int\left[\frac{1}{\pi} \left(\frac{d m_{23}^2}{(m_{23}^2-m_C^2)^2}\right)\right]
 {\rm Tr} \left[ 
 [\tilde{\Gamma}_1(A\rightarrow B C^*)]
 \tilde{\Gamma}_2 (C^*\rightarrow D E)  \right].
\end{eqnarray}
Here, $k^2 \equiv m_{23}^2$, $f_s$ is the symmetry factor, and $m_i$ $(i=A,B,C,D,E)$ is the mass of 
the $i^{th}$ particle. 

The width functions are
\begin{eqnarray}
[\tilde{\Gamma}_1]^i_j &=& \int \frac{d_{PS}^{A\rightarrow B C^*}}{2}
\left[|\mathcal{M}_1(A \rightarrow B C^*)|^2\right]^i_j 
=\frac{1}{2}\int \frac{\bar{\beta}}{8\pi}\frac{d \cos\theta}{2}\frac{d\phi}{2\pi}
\left[|\mathcal{M}_1(A\rightarrow B C^*)|^2\right]^i_j, \nonumber\\
 {[\tilde{\Gamma}_2]}^j_i &=& \int \frac{d_{PS}^{C^*\rightarrow D E}}{2}
 \left[|\mathcal{M}_2(C^* \rightarrow D E)|^2\right]^j_i 
  =\frac{1}{2}\int \frac{\bar{\beta}_{23}}{8\pi}\frac{d \cos\theta_{23}}{2}\frac{d\phi_{23}}{2\pi}
  \left[|\mathcal{M}_2(C^*\rightarrow D E)|^2\right]^j_i\,,
\end{eqnarray}
where the boost factors are
\begin{eqnarray}
\label{3_dec_callan_1}
\bar{\beta} &=& \sqrt{1-\frac{2(m_B^2+m_{23}^2)}{m_A^2}+\frac{(m_B^2-m_{23}^2)^2}{m_A^4}},\\
\bar{\beta}_{23} &=& \sqrt{1-\frac{2(m_D^2+m_E^2)}{m_{23}^2}+\frac{(m_D^2-m_E^2)^2}{m_{23}^4}}.
\end{eqnarray}
The indices $i$ and $j$ are Lorentz or spin indices for spin-1 or spin-$\frac12$ particles respectively.
For scalar propagators there are no such indices, the respective $\tilde{\Gamma}$ is a number 
rather than a matrix.

As the phase space measure $d^3{\bf p}/2E$ is Lorentz invariant, every two-body phase space can be 
computed in a reference frame where the decaying particle is considered to be at rest. 
These subspaces are to be joined using proper boost factors.

Considering $A$ to be at  rest, the four-momenta of $B$ and $C^*$ can be written as:
\begin{align}
p_1 = & \frac{m_A}{2}\left( 1 + \frac{m_B^2}{m_A^2} - \frac{m_{23}^2}{m_A^2}, 0 ,0 ,-\bar{\beta}\right), \\
k = & \frac{m_A}{2}\left( 1 - \frac{m_B^2}{m_A^2} + \frac{m_{23}^2}{m_A^2}, 0 ,0 ,\bar{\beta}\right).
\end{align}
The boost factor from the rest frame of $C^*$ towards the rest frame of $A$ is
\begin{align}
\gamma  \ =  \ \frac{k^0}{\sqrt{m_{23}^2}} = \frac{m_A}{2 \sqrt{m_{23}^2}}\left( 1 - \frac{m_B^2}{m_A^2} + \frac{m_{23}^2}{m_A^2}\right), \; \;
\gamma\beta \ =  \ \frac{m_A}{2 \sqrt{m_{23}^2}}\bar{\beta}.
\end{align}

Similarly, in  the rest frame of $C^*$, the four-momenta of  $D$ and $E$  are
\begin{align}
\hat{p}_2 = & \ \frac{\sqrt{m_{23}^2}}{2}\left( 1 + \frac{m_D^2}{m_{23}^2} - \frac{m_E^2}{m_{23}^2} , \bar{\beta}_{23} \sin\theta_{23} , 0 , \bar{\beta}_{23}\cos\theta_{23}\right), \\
\hat{p}_3 = & \ \frac{\sqrt{m_{23}^2}}{2}\left( 1 - \frac{m_D^2}{m_{23}^2} + \frac{m_E^2}{m_{23}^2} , - \bar{\beta}_{23} \sin\theta_{23} , 0 , - \bar{\beta}_{23}\cos\theta_{23}\right)\,,
\end{align}
and hence in the rest frame of $A$ they are 
\begin{align}
p_2 = & \ \frac{\sqrt{m_{23}^2}}{2}
\begin{pmatrix}
 \gamma \left(1 + \frac{m_D^2}{m_{23}^2} - \frac{m_E^2}{m_{23}^2} \right) + \gamma\beta \bar{\beta}_{23}\cos\theta_{23} \\ \bar{\beta}_{23} \sin\theta_{23} \\ 0 \\ \gamma\beta \left(1 + \frac{m_D^2}{m_{23}^2} - \frac{m_E^2}{m_{23}^2} \right) + \gamma\bar{\beta}_{23}\cos\theta_{23} 
\end{pmatrix}, \\ \nonumber \\
p_3 = & \ \frac{\sqrt{m_{23}^2}}{2}
\begin{pmatrix}
 \gamma \left(1 - \frac{m_D^2}{m_{23}^2} + \frac{m_E^2}{m_{23}^2} \right) - \gamma\beta \bar{\beta}_{23}\cos\theta_{23} \\ - \bar{\beta}_{23} \sin\theta_{23} \\ 0 \\ \gamma\beta \left(1 - \frac{m_D^2}{m_{23}^2} + \frac{m_E^2}{m_{23}^2} \right) - \gamma\bar{\beta}_{23}\cos\theta_{23} 
\end{pmatrix}.
\end{align}

This is sufficient to perform the phase space integration.

\subsection{Four-body decay} 

For the decay $A(p) \to 1(p_1) 2(p_2) 3(p_3) 4(p_4)$, 
let us consider both symmetric and asymmetric cascades.

\subsubsection{Symmetric Decay} \label{Sym_Decay}

\begin{figure}[h!]
\begin{center}
\epsfig{figure=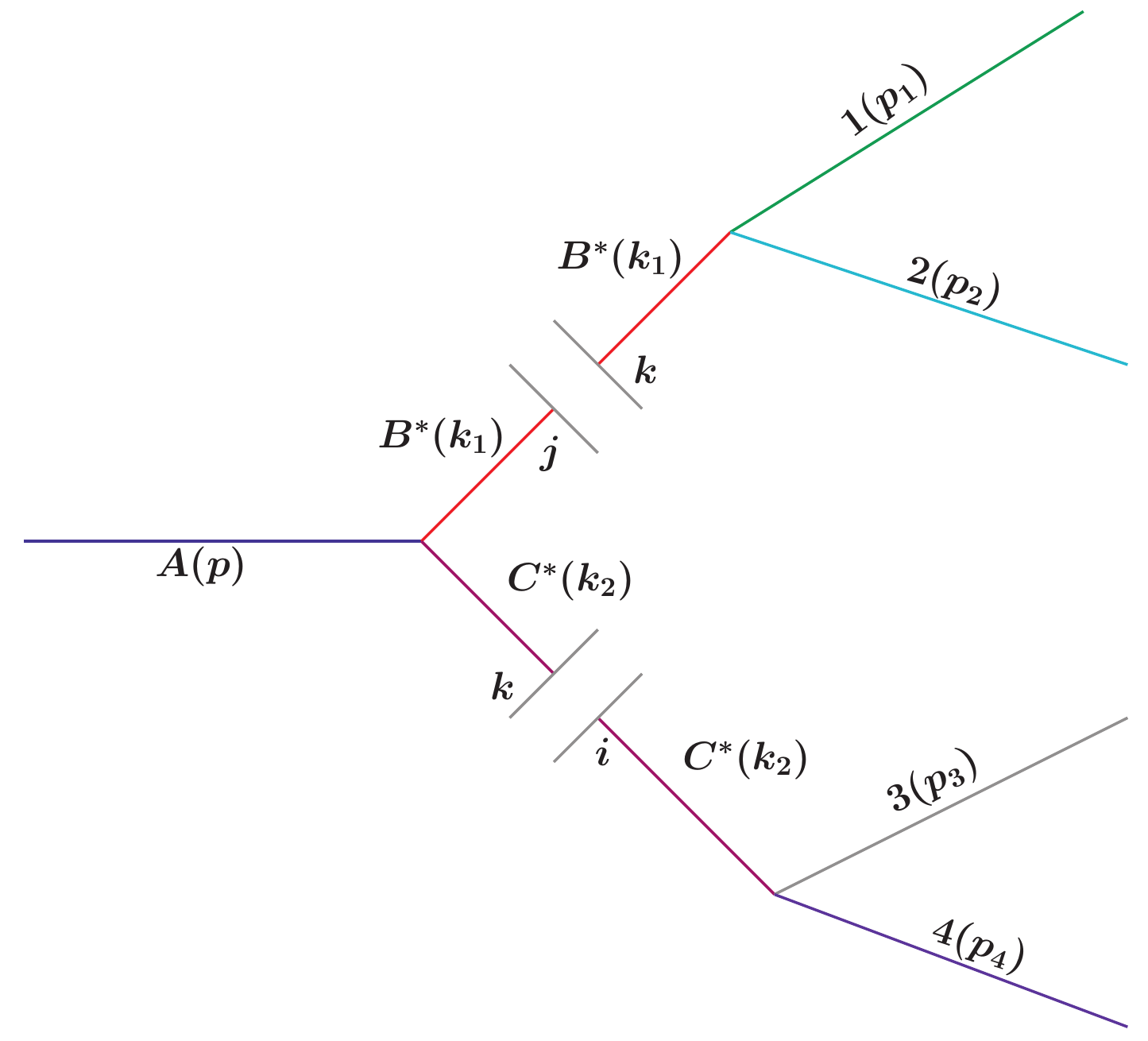,scale=0.575}
\end{center}
\caption{\label{fig 2.1}  Representative diagram for the symmetric four-body cascade
$A\rightarrow 1234$.
}
\end{figure}

To start with let us first discuss the symmetric decay chain leading to four-body final state.
 Let us consider decay of the particle $A$ as: 
$A(p)\rightarrow B^*(q_{12}) C^*(q_{34})$, followed by decays of off-shell propagators:  $B^*(q_{12}) \rightarrow 1 (p_1) \; 2(p_2) , C^* \rightarrow 3(p_3) \; 4(p_4).$

Using our prescription decay width can be written as:
\begin{eqnarray}
\Gamma(A\rightarrow 1 2 3 4)&=&\frac{1}{m_A}\int\left[\frac{1}{\pi} \left(\frac{d m_{12}^2}{(m_{12}^2-m_{B}^2)^2}\right)\right]
\int\left[\frac{1}{\pi} \left(\frac{d m_{34}^2}{(m_{34}^2-m_{C}^2)^2}\right)\right]
 f_s \nonumber
 \\&&
 {\rm Tr} \left[ 
 [\tilde{\Gamma}_1(A\rightarrow B^* C^*)]
 \tilde{\Gamma}_2 (B^*\rightarrow 1 2)\tilde{\Gamma}_3 (C^*\rightarrow 3 4)  \right] .
\end{eqnarray}

The $\tilde{\Gamma}$ functions are expressed as:
\begin{align}
\bigg[\tilde{\Gamma}_1(A\rightarrow B^* C^*)\bigg]^i_j \ = & \ \int \frac{d_{PS}^{A\rightarrow B^* C^*}}{2}\bigg[|\mathcal{M}_1(A \rightarrow B^* C^*)|^2\bigg]^i_j ,\nonumber
\\ \
= & \frac{1}{2}\int \frac{\bar{\beta}}{8\pi}\frac{d \cos\theta}{2}\frac{d\phi}{2\pi}
[|\mathcal{M}_1(A\rightarrow B^* C^*)|^2]^i_j
,\\
\bigg[\tilde{\Gamma}_2(B^*\rightarrow 1 2)\bigg]^j_k \ = & \ \int \frac{d_{PS}^{B^*\rightarrow 1 2}}{2}\bigg[|\mathcal{M}_2(B^* \rightarrow 1 2)|^2\bigg]^j_k ,\nonumber
\\ \
= & \frac{1}{2}\int \frac{\bar{\beta}_{12}}{8\pi}\frac{d \cos\theta_{12}}{2}\frac{d\phi_{12}}{2\pi}
[|\mathcal{M}_2(B^*\rightarrow 1 2)|^2]^j_k
,\\
\bigg[\tilde{\Gamma}_3 (C^*\rightarrow 3 4)\bigg]^k_i \ = & \ \int \frac{d_{PS}^{C^*\rightarrow 3 4}}{2}\bigg[|\mathcal{M}_3(C^* \rightarrow 3 4)|^2\bigg]^k_i ,\nonumber
\\ \
= & \frac{1}{2}\int \frac{\bar{\beta}_{34}}{8\pi}\frac{d \cos\theta_{34}}{2}\frac{d\phi_{34}}{2\pi}
[|\mathcal{M}_3(C^*\rightarrow 3 4)|^2]^k_i,
\end{align}
where,
\begin{align}
\bar{\beta}\left(\frac{m_{12}^2}{m_A^2}, \frac{m_{34}^2}{m_A^2}\right) \ = & \ \sqrt{1 - \frac{2\left(m_{12}^2 + m_{34}^2\right)}{m_A^2} + \frac{\left( m_{12}^2 - m_{34}^2\right)^2}{m_A^4}} ,\\
\bar{\beta}_{12}\left(\frac{m_{1}^2}{m_{12}^2}, \frac{m_{2}^2}{m_{12}^2}\right) \ = & \ \sqrt{1 - \frac{2\left(m_{1}^2 + m_{2}^2\right)}{m_{12}^2} + \frac{\left( m_{1}^2 - m_{2}^2\right)^2}{m_{12}^4}}, \\
\bar{\beta}_{34}\left(\frac{m_{3}^2}{m_{34}^2}, \frac{m_{4}^2}{m_{34}^2}\right) \ = & \ \sqrt{1 - \frac{2\left(m_{3}^2 + m_{4}^2\right)}{m_{34}^2} + \frac{\left( m_{3}^2 - m_{4}^2\right)^2}{m_{34}^4}}.
\end{align}

Combining all the contributions, we find
\begin{eqnarray}
\Gamma(A\rightarrow 1 2 3 4)&=&\frac{f_s}{m_A}\int\left[\frac{1}{\pi} \left(\frac{d m_{12}^2}{(m_{12}^2-m_{B}^2)^2}\right)\right]
\int\left[\frac{1}{\pi} \left(\frac{d m_{34}^2}{(m_{34}^2-m_{C}^2)^2}\right)\right]
  \nonumber
 \\&&
 \frac{1}{2^3}\int \frac{\bar{\beta}}{8\pi}\frac{d \cos\theta}{2}\frac{d\phi}{2\pi}\int \frac{\bar{\beta}_{12}}{8\pi}\frac{d \cos\theta_{12}}{2}\frac{d\phi_{12}}{2\pi}\int \frac{\bar{\beta}_{34}}{8\pi}\frac{d \cos\theta_{34}}{2}\frac{d\phi_{34}}{2\pi}
 \\&&
 \left[ 
 [|\mathcal{M}_1(A\rightarrow B^* C^*)|^2]^i_j
 [|\mathcal{M}_2 (B^*\rightarrow 1 2)|^2]^j_k [|\mathcal{M}_3 (C^*\rightarrow 3 4)|^2]^k_i  \right].\nonumber
\end{eqnarray}

In the centre-of-mass (CM) frame we have following momenta:
\begin{align}
q_{12} \ = & \ \frac{m_A}{2}\left( 1 + \frac{m_{12}^2}{m_A^2} - \frac{m_{34}^2}{m_A^2} , 0 , 0 , \bar{\beta} \right), \label{q12}\\
q_{34} \ = & \ \frac{m_A}{2}\left( 1 - \frac{m_{12}^2}{m_A^2} + \frac{m_{34}^2}{m_A^2} , 0 , 0 , - \bar{\beta} \right).\label{q34}
\end{align}

The necessary boost factors from the rest frame of $q_{12}$ towards the CM frame can be written as:
\begin{align}
\gamma_1 \ = & \ \frac{q_{12}^0}{\sqrt{m_{12}^2}} \ = 
\ \frac{m_A}{2 \sqrt{m_{12}^2}} \left( 1 + \frac{m_{12}^2}{m_A^2} - \frac{m_{34}^2}{m_A^2} \right), \label{2_1_1} \\
\gamma_1 \beta_1 \ = & \ \frac{q_{12}^0}{\sqrt{m_{12}^2}}.\frac{|\bm{p}_{12}|}{q_{12}^0} \ = \ \frac{m_A}{2 \sqrt{m_{12}^2}} \bar{\beta}. \label{2_1_2}
\end{align}

One can similarly write the same from the rest frame of $q_{34}$ as:
\begin{align}
\gamma_2 \ = & \ \frac{q_{34}^0}{\sqrt{m_{34}^2}} \ = 
\ \frac{m_A}{2 \sqrt{m_{34}^2}} \left( 1 - \frac{m_{12}^2}{m_A^2} + \frac{m_{34}^2}{m_A^2} \right) ,\label{2_1_3}\\
\gamma_2 \beta_2 \ = & \ \frac{q_{34}^0}{\sqrt{m_{34}^2}}.\frac{|\bm{p}_{34}|}{q_{34}^0} \ = 
\ \frac{m_A}{2 \sqrt{m_{34}^2}} \bar{\beta}.\label{2_1_4}
\end{align}

In the rest frame of $q_{12}$ the momenta components of the particles $1$ and $2$  can be given as:
\begin{align}
\hat{p}_1 \ = & \ \frac{\sqrt{m_{12}^2}}{2}\left( 1 + \frac{m_1^2}{m_{12}^2} - \frac{m_2^2}{m_{12}^2}, \bar{\beta}_{12} \sin\theta_{12} , 0 , \bar{\beta}_{12}\cos\theta_{12} \right), \\
\hat{p}_2 \ = & \ \frac{\sqrt{m_{12}^2}}{2}\left( 1 - \frac{m_1^2}{m_{12}^2} + \frac{m_2^2}{m_{12}^2}, -\bar{\beta}_{12} \sin\theta_{12} , 0 , - \bar{\beta}_{12}\cos\theta_{12} \right).
\end{align}

The momenta of $3$ and $4$,  in the rest frame of $q_{34}$, can be given as:
\begin{align}
\hat{p}_3 \ = & \ \frac{\sqrt{m_{34}^2}}{2}\left( 1 + \frac{m_3^2}{m_{34}^2} - \frac{m_4^2}{m_{34}^2}, \bar{\beta}_{34} \sin\theta_{34}\cos\phi_{34} , \bar{\beta}_{34} \sin\theta_{34}\sin\phi_{34} , \bar{\beta}_{34}\cos\theta_{34} \right) ,\\
\hat{p}_4 \ = & \ \frac{\sqrt{m_{34}^2}}{2}\left( 1 - \frac{m_3^2}{m_{34}^2} + \frac{m_4^2}{m_{34}^2}, -\bar{\beta}_{34} \sin\theta_{34}\cos\phi_{34} , - \bar{\beta}_{34} \sin\theta_{34}\sin\phi_{34} , - \bar{\beta}_{34}\cos\theta_{34} \right).
\end{align}

Thus after including boost factors one can write down the  momenta of  $1$, $2$, $3$ and $4$, in the CM
frame, as :
\begin{align}
p_1 = & \frac{\sqrt{m_{12}^2}}{2}
\begin{pmatrix}
\gamma_1 \left( 1 + \frac{m_1^2}{m_{12}^2} - \frac{m_2^2}{m_{12}^2}\right) +\gamma_1 \beta_1 \bar{\beta}_{12}\cos\theta_{12} \\
\bar{\beta}_{12}\sin\theta_{12} \\
0 \\
\gamma_1 \beta_1 \left( 1 + \frac{m_1^2}{m_{12}^2} - \frac{m_2^2}{m_{12}^2}\right) +\gamma_1  \bar{\beta}_{12}\cos\theta_{12}
\end{pmatrix},\label{sym_p_1}
\\ \nonumber \\
p_2 = & \frac{\sqrt{m_{12}^2}}{2}
\begin{pmatrix}
\gamma_1 \left( 1 - \frac{m_1^2}{m_{12}^2} + \frac{m_2^2}{m_{12}^2}\right) - \gamma_1 \beta_1 \bar{\beta}_{12}\cos\theta_{12} \\
-\bar{\beta}_{12}\sin\theta_{12} \\
0 \\
\gamma_1 \beta_1 \left( 1 - \frac{m_1^2}{m_{12}^2} + \frac{m_2^2}{m_{12}^2}\right) - \gamma_1  \bar{\beta}_{12}\cos\theta_{12}
\end{pmatrix},\label{sym_p_2}
\\ \nonumber \\
p_3 = & \frac{\sqrt{m_{34}^2}}{2}
\begin{pmatrix}
\gamma_2\left( 1 + \frac{m_3^2}{m_{34}^2} - \frac{m_4^2}{m_{34}^2}\right) + \gamma_2 \beta_2 \bar{\beta}_{34}\cos\theta_{34} \\
\bar{\beta}_{34} \sin\theta_{34}\cos\phi_{34} \\
\bar{\beta}_{34} \sin\theta_{34}\sin\phi_{34} \\
\gamma_2 \beta_2\left( 1 + \frac{m_3^2}{m_{34}^2} - \frac{m_4^2}{m_{34}^2}\right) + \gamma_2 \bar{\beta}_{34}\cos\theta_{34}
\end{pmatrix},\label{sym_p_3}
\\ \nonumber \\
p_4 = & \frac{\sqrt{m_{34}^2}}{2}
\begin{pmatrix}
\gamma_2\left( 1 - \frac{m_3^2}{m_{34}^2} + \frac{m_4^2}{m_{34}^2}\right) - \gamma_2 \beta_2 \bar{\beta}_{34}\cos\theta_{34} \\
- \bar{\beta}_{34} \sin\theta_{34}\cos\phi_{34} \\
-\bar{\beta}_{34} \sin\theta_{34}\sin\phi_{34} \\
\gamma_2 \beta_2\left( 1 - \frac{m_3^2}{m_{34}^2} + \frac{m_4^2}{m_{34}^2}\right) - \gamma_2 \bar{\beta}_{34}\cos\theta_{34}
\end{pmatrix},\label{sym_p_4}
\end{align}
respectively.

These details are sufficient to perform the phase space integration and to compute the decay width for the full cascade.

 
\subsubsection{Asymmetric Decay}\label{Asym_Decay}
\begin{figure}[h!]
\begin{center}
\epsfig{figure=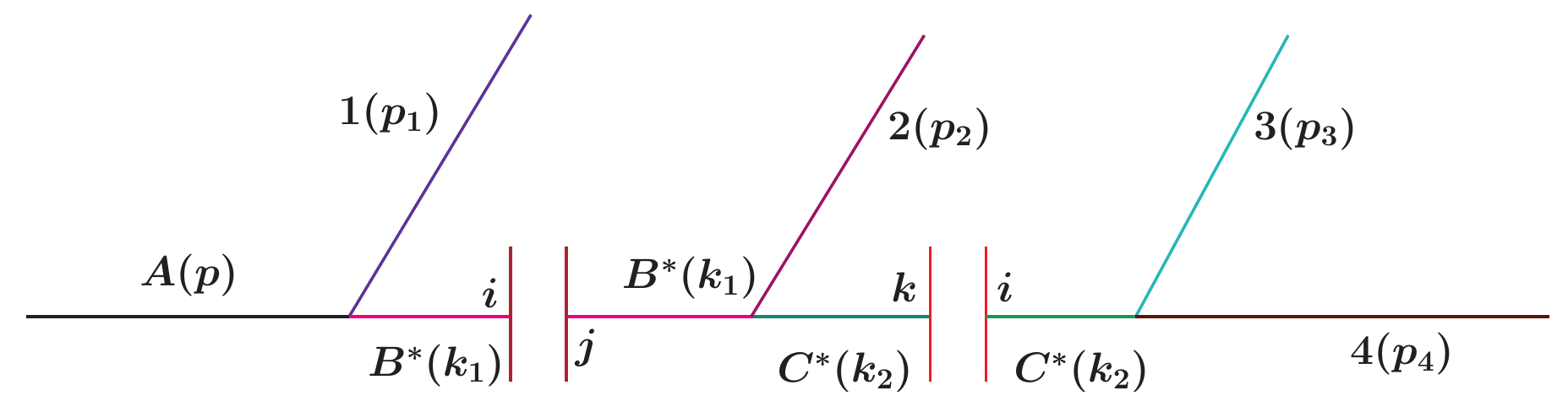,scale=0.575}
\end{center}
\caption{\label{fig1.2} Representative diagram for the asymmetric four-body cascade  
$A\rightarrow 123 4 $.
}
\end{figure}

Let us consider the identical decay $A\to 1234$ here but with a different cascade topology; this is 
an asymmetric cascade (Fig.~\ref{fig1.2}), as both the off-shell propagators are appearing in the same chain: 
$A(p)\rightarrow 1(p_1) B^* (k_1), \ B^*(k_1) \rightarrow 2(p_2) C^*(k_2), \ C^* (k_2) \rightarrow 3(p_3) 4(p_4)$.

According to our proposal, we write the decay width as 
\begin{eqnarray}
\Gamma(A\rightarrow 1 2 3 4)&=&\frac{f_s}{m_A}\int\left[\frac{1}{\pi} \left(\frac{d m_{12}^2}{(m_{12}^2-m_{B}^2)^2}\right)\right]
\int\left[\frac{1}{\pi} \left(\frac{d m_{23}^2}{(m_{23}^2-m_{C}^2)^2}\right)\right]
  \nonumber
 \\&&
 \times
  {\rm Tr} \left[ 
 [\tilde{\Gamma}_1(A\rightarrow B^* 1)]
 \tilde{\Gamma}_2 (B^*\rightarrow C^* 2)\tilde{\Gamma}_3 (C^*\rightarrow 3 4)  \right] ,
\end{eqnarray}
where the $\tilde{\Gamma}$ functions are given as:
\begin{align}
\bigg[\tilde{\Gamma}_1(A\rightarrow B^* 1)\bigg]^i_j \ = & \ \int \frac{d_{PS}^{A\rightarrow B^* 1}}{2}\bigg[|\mathcal{M}_1(A \rightarrow B^* 1)|^2\bigg]^i_j \nonumber
\\ \
= & \frac{1}{2}\int \frac{\bar{\beta}_1}{8\pi}\frac{d \cos\theta}{2}\frac{d\phi}{2\pi}
[|\mathcal{M}_1(A\rightarrow B^* 1)|^2]^i_j,
\\
\bigg[\tilde{\Gamma}_2(B^*\rightarrow C^* 2)\bigg]^j_k \ = & \ \int \frac{d_{PS}^{B^*\rightarrow C^* 2}}{2}\bigg[|\mathcal{M}_2(B^* \rightarrow C^* 2)|^2\bigg]^j_k\nonumber
\\ \
= & \frac{1}{2}\int \frac{\bar{\beta}_{2}}{8\pi}\frac{d \cos\theta_{1}}{2}\frac{d\phi_{1}}{2\pi}
[|\mathcal{M}_2(B^*\rightarrow C^* 2)|^2]^j_k
,\\
\bigg[\tilde{\Gamma}_3 (C^*\rightarrow 3 4)\bigg]^k_i \ = & \ \int \frac{d_{PS}^{C^*\rightarrow 3 4}}{2}\bigg[|\mathcal{M}_3(C^* \rightarrow 3 4)|^2\bigg]^k_i \nonumber
\\ \
= & \frac{1}{2}\int \frac{\bar{\beta}_{3}}{8\pi}\frac{d \cos\theta_{2}}{2}\frac{d\phi_{2}}{2\pi}
[|\mathcal{M}_3(C^*\rightarrow 3 4)|^2]^k_i.
\end{align}

The boost factors are written as:
\begin{align}
\bar{\beta}_1\left(\frac{m_{1}^2}{m_A^2}, \frac{m_{12}^2}{m_A^2}\right) \ = & \ \sqrt{1 - \frac{2\left(m_{1}^2 + m_{12}^2\right)}{m_A^2} + \frac{\left( m_{1}^2 - m_{12}^2\right)^2}{m_A^4}}, \\
\bar{\beta}_{2}\left(\frac{m_{2}^2}{m_{12}^2}, \frac{m_{23}^2}{m_{12}^2}\right) \ = & \ \sqrt{1 - \frac{2\left(m_{2}^2 + m_{23}^2\right)}{m_{12}^2} + \frac{\left( m_{2}^2 - m_{23}^2\right)^2}{m_{12}^4}} ,\\
\bar{\beta}_{3}\left(\frac{m_{3}^2}{m_{23}^2}, \frac{m_{4}^2}{m_{23}^2}\right) \ = & \ \sqrt{1 - \frac{2\left(m_{3}^2 + m_{4}^2\right)}{m_{23}^2} + \frac{\left( m_{3}^2 - m_{4}^2\right)^2}{m_{23}^4}}.
\end{align}
\begin{figure}[h!]
\begin{center}
\includegraphics[trim=3cm 0 0 5cm,clip,scale=0.7]{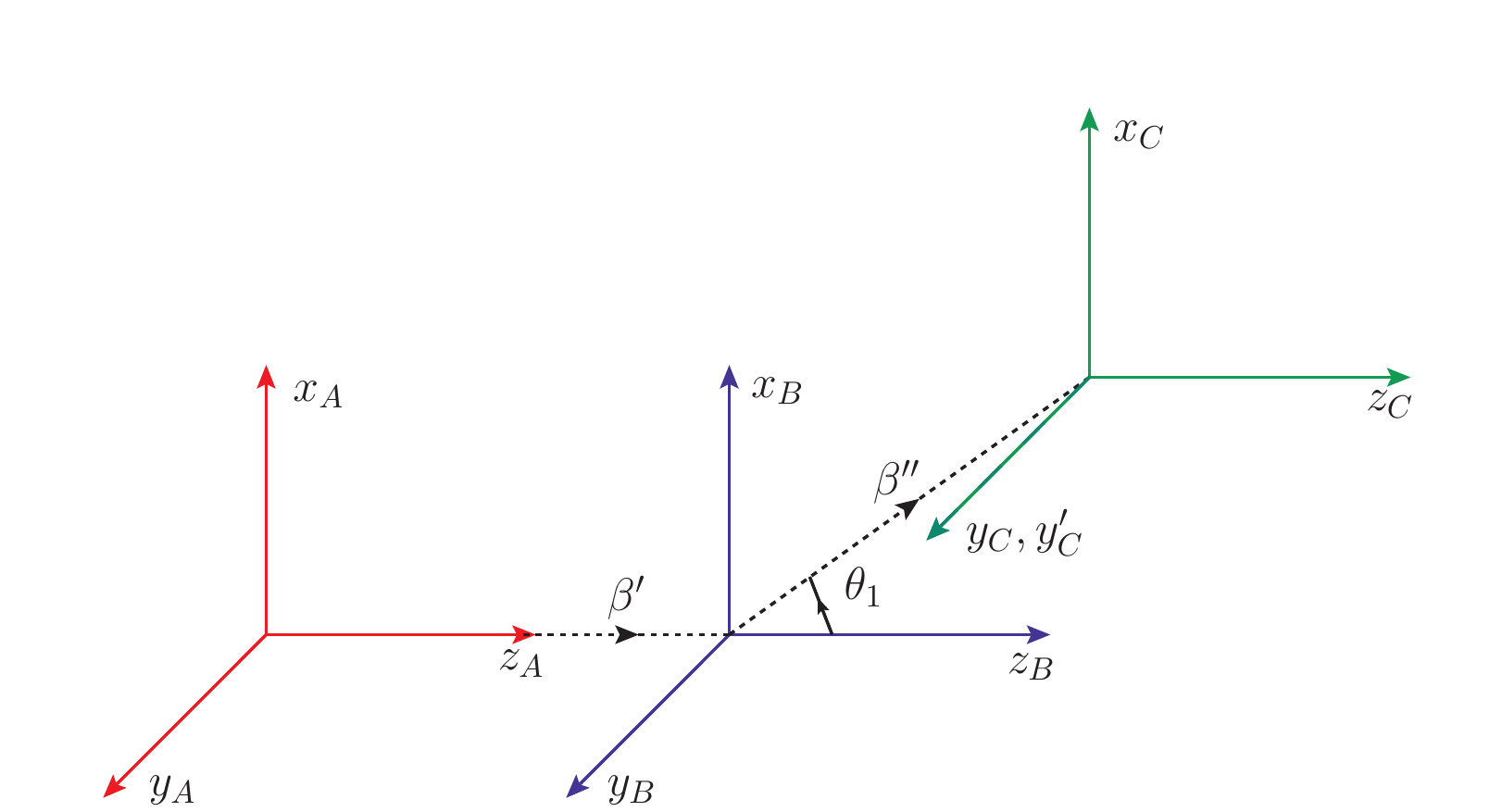}
\end{center}
\caption{\label{fig2.3}  A schematic diagram of rest frames of $A$, $B^*$ and $C^*$.
}
\end{figure}

To explain the phase space structure we define $S_A$, $S_B$ and $S_C$ to be  the rest frames of $A$, $B^*$ and $C^*$ respectively, see Fig.~\ref{fig2.3}.

In $S_A$, the components of momenta of  $1$ and $B^*$ are given as:
\begin{align}
p_1 \ = & \ \frac{m_A}{2}\left(1 + \frac{m_1^2}{m_A^2} - \frac{m_{12}^2}{m_A^2} , 0, 0, -\bar{\beta}_1 \right), \label{e_2.2.1} \\
k_1 \ = & \ \frac{m_A}{2}\left(1 - \frac{m_1^2}{m_A^2} + \frac{m_{12}^2}{m_A^2} , 0, 0, \bar{\beta}_1 \right). \label{e_2.2.2}
\end{align}

The same for  $2$ and $C^*$ in $S_B$ are given as:
\begin{align}
\hat{p}_2 \ = & \frac{\sqrt{m_{12}^2}}{2} \left( 1 + \frac{m_2^2}{m_{12}^2} - \frac{m_{23}^2}{m_{12}^2} , -\bar{\beta}_2\sin\theta_1, 0, -\bar{\beta}_2\cos\theta_1 \right), \\
\hat{k}_2 \ = & \frac{\sqrt{m_{12}^2}}{2} \left( 1 - \frac{m_2^2}{m_{12}^2} + \frac{m_{23}^2}{m_{12}^2} , \bar{\beta}_2\sin\theta_1, 0, \bar{\beta}_2\cos\theta_1 \right),
\end{align}
and similarly for  $3$ and $4$ in $S_C$ the momenta are
\begin{align}
\hat{p}_3 \ = & \frac{\sqrt{m_{23}^2}}{2} \left( 1 + \frac{m_3^2}{m_{23}^2} - \frac{m_{4}^2}{m_{23}^2} , \bar{\beta}_3\sin\theta_2\cos\phi_2, \bar{\beta}_3\sin\theta_2\sin\phi_2, \bar{\beta}_3\cos\theta_2 \right), \\
\hat{p}_4 \ = & \frac{\sqrt{m_{23}^2}}{2} \left( 1 - \frac{m_3^2}{m_{23}^2} + \frac{m_{4}^2}{m_{23}^2} , -\bar{\beta}_3\sin\theta_2\cos\phi_2,- \bar{\beta}_3\sin\theta_2\sin\phi_2, -\bar{\beta}_3\cos\theta_2 \right).
\end{align}

The necessary boost factors from  $S_A$ to $S_B$ are
\begin{align}
\bm{\beta'}  =  \frac{|\bm{k_1}|}{k_1^0}\hat{z}, \ \qquad \gamma ' = \frac{k_1^0}{\sqrt{m_{12}^2}},  \ \qquad
\gamma ' \beta '  = \frac{m_A}{2}\frac{\bar{\beta}_1}{\sqrt{m_{12}^2}},
\end{align}
and the  Lorentz transformation matrix from $S_B$ to $S_A$ is
\begin{align}
\Lambda ' = 
\begin{pmatrix}
\gamma ' & 0 & 0 & \gamma '\beta ' \\
0 & 1 & 0 & 0 \\
0 & 0 & 1 & 0 \\
\gamma '\beta ' & 0 & 0 & \gamma '
\end{pmatrix}.
\end{align}

Thus the  boost factors from $S_B$ to $S_C$ are
\begin{align}
\bm{\beta ''} =  \frac{|\bm{k_2}|}{k_2^0}\left( \sin\theta_1 \hat{x} + \cos\theta_1 \hat{z} \right), \ \qquad
\gamma '' \beta '' =  \frac{\sqrt{m_{12}^2}}{2}\frac{\bar{\beta}_2}{\sqrt{m_{23}^2}}.
\end{align}

Using the velocity addition theorem we can write down  the components of the boost from $S_A$ to $S_C$ as:
\begin{align}
\beta_x = & \frac{\beta_x ''}{\gamma ' (1 + \beta ' \beta_z '')} = \frac{\frac{\sqrt{m_{12}^2}}{k_1^0}
\frac{|\bm{k_2}|}{k_2^0}\sin\theta_1}{1 + \frac{|\bm{k_1}|}{k_1^0}\, \frac{|\bm{k_2}|}{k_2^0}\cos\theta_1}\,,\ \ 
\beta_y =  \frac{\beta_y ''}{\gamma ' (1 + \beta ' \beta_z '')} = 0\,,\nonumber\\
\beta_z = & \frac{\beta ' + \beta_z ''}{1 + \beta ' \beta_z ''} = \frac{\frac{|\bm{k_1}|}{k_1^0} + \frac{|\bm{k_2}|}{k_2^0}
\cos\theta_1 }{1 + \frac{|\bm{k_1}|}{k_1^0}\, \frac{|\bm{k_2}|}{k_2^0}\cos\theta_1}\,,\ \ 
\gamma =  \ \frac{1}{\sqrt{1-\beta_x^2-\beta_z^2}}.
\end{align}

Now combining all these we can write down the Lorentz transformation matrix from $S_C$ to $S_A$ :
\begin{equation}
\Lambda = 
\begin{pmatrix}
\gamma & \gamma \beta_x & 0 & \gamma \beta_z \\
\gamma \beta_x & 1+(\gamma-1)\frac{\beta_x^2}{\beta^2} & 0 & (\gamma - 1)\frac{\beta_x \beta_z}{\beta^2} \\
0 & 0 & 1 & 0 \\
\gamma \beta_z & (\gamma - 1)\frac{\beta_x \beta_z}{\beta^2} & 0 & 1+(\gamma-1)\frac{\beta_z^2}{\beta^2}
\end{pmatrix}.
\end{equation}

This helps us to compute the momenta of $2$ and $C^*$ in the rest frame of $A$, i.e., $S_A$,  as :
\begin{align}
p_2 = \frac{\gamma ' \sqrt{m_{12}^2}}{2}
\begin{pmatrix}
\gamma ' \left( 1 + \frac{m_2^2}{m_{12}^2} - \frac{m_{23}^2}{m_{12}^2} \right) - \gamma ' \beta ' \bar{\beta}_2\cos\theta_1 \\
-\bar{\beta}_2\sin\theta_1 \\
0 \\
\gamma ' \beta ' \left( 1 + \frac{m_2^2}{m_{12}^2} - \frac{m_{23}^2}{m_{12}^2} \right) -\gamma ' \bar{\beta}_2\cos\theta_1
\end{pmatrix}, \label{e_2.2.3} \\ \nonumber \\
k_2 = \frac{\gamma ' \sqrt{m_{12}^2}}{2}
\begin{pmatrix}
\gamma ' \left( 1 - \frac{m_2^2}{m_{12}^2} + \frac{m_{23}^2}{m_{12}^2} \right) + \gamma ' \beta ' \bar{\beta}_2\cos\theta_1  \\
\bar{\beta}_2\sin\theta_1 \\
0 \\
\gamma ' \beta ' \left( 1 - \frac{m_2^2}{m_{12}^2} + \frac{m_{23}^2}{m_{12}^2} \right) +\gamma ' \bar{\beta}_2\cos\theta_1
\end{pmatrix}, \label{e_2.2.4}
\end{align}
and, that for  $3$ and $4$ as:
\begin{align}
p_3 =  \frac{\sqrt{m_{23}^2}}{2}
\begin{pmatrix}
 \gamma\Big[ \left( 1 + \frac{m_3^2}{m_{23}^2} - \frac{m_{4}^2}{m_{23}^2}\right) + \beta_x\bar{\beta}_3 \sin\theta_2\cos\phi_2 + \beta_z\bar{\beta}_3 \cos\theta_2 \Big] \\
\gamma\beta_x \left(1 + \frac{m_3^2}{m_{23}^2} - \frac{m_{4}^2}{m_{23}^2}\right) + \Big\{ 1+(\gamma-1)\frac{\beta_x^2}{\beta^2}\Big\} \bar{\beta}_3 \sin\theta_2\cos\phi_2 + (\gamma - 1)\frac{\beta_x \beta_z}{\beta^2}\bar{\beta}_3 \cos\theta_2 \\
 \bar{\beta}_3\sin \theta_2 \sin\phi_2  \\ 
\gamma\beta_z\left(1 + \frac{m_3^2}{m_{23}^2} - \frac{m_{4}^2}{m_{23}^2}\right) + (\gamma - 1)\frac{\beta_x \beta_z}{\beta^2} \bar{\beta}_3 \sin\theta_2\cos\phi_2 + \Big\{ 1+(\gamma-1)\frac{\beta_z^2}{\beta^2}\Big\} \bar{\beta}_3 \cos\theta_2
\end{pmatrix}, \label{e_2.2.5}
\\ \nonumber \\ 
p_4 = \frac{\sqrt{m_{23}^2}}{2}
\begin{pmatrix}
 \gamma\Big[ \left( 1 - \frac{m_3^2}{m_{23}^2} + \frac{m_{4}^2}{m_{23}^2}\right) - \beta_x\bar{\beta}_3 \sin\theta_2\cos\phi_2 - \beta_z\bar{\beta}_3 \cos\theta_2 \Big] \\
\gamma\beta_x \left(1 - \frac{m_3^2}{m_{23}^2} + \frac{m_{4}^2}{m_{23}^2}\right) - \Big\{ 1+(\gamma-1)\frac{\beta_x^2}{\beta^2}\Big\} \bar{\beta}_3 \sin\theta_2\cos\phi_2 - (\gamma - 1)\frac{\beta_x \beta_z}{\beta^2}\bar{\beta}_3 \cos\theta_2 \\
- \bar{\beta}_3\sin \theta_2 \sin\phi_2  \\ 
\gamma\beta_z\left(1 - \frac{m_3^2}{m_{23}^2} + \frac{m_{4}^2}{m_{23}^2}\right) - (\gamma - 1)\frac{\beta_x \beta_z}{\beta^2} \bar{\beta}_3 \sin\theta_2\cos\phi_2 - \Big\{ 1+(\gamma-1)\frac{\beta_z^2}{\beta^2}\Big\} \bar{\beta}_3 \cos\theta_2
\end{pmatrix}, \label{e_2.2.6}
\end{align}
respectively. One can now compute the integration numerically using all these momenta in the $S_A$ frame.

\subsection{Interference of amplitudes of four-body decay through symmetric and asymmetric channels}

So far we have discussed the solitary contributions to the four body phase space from either symmetric or asymmetric cascade. We have mentioned repeated that it is indeed possible to have more than one cascade decay chain leading to same final states.  Now if both the cascades  are either symmetric or asymmetric then the phase space for the interference diagrams would be trivial. If one of them is symmetric and other one is asymmetric, then we can have two possible structures depending on the positions of the external particles: (i) same for both cascades, (ii) shuffled among themselves. We have discussed the earlier case in detail in text.

\begin{figure}[h!]
\begin{center}
\epsfig{figure=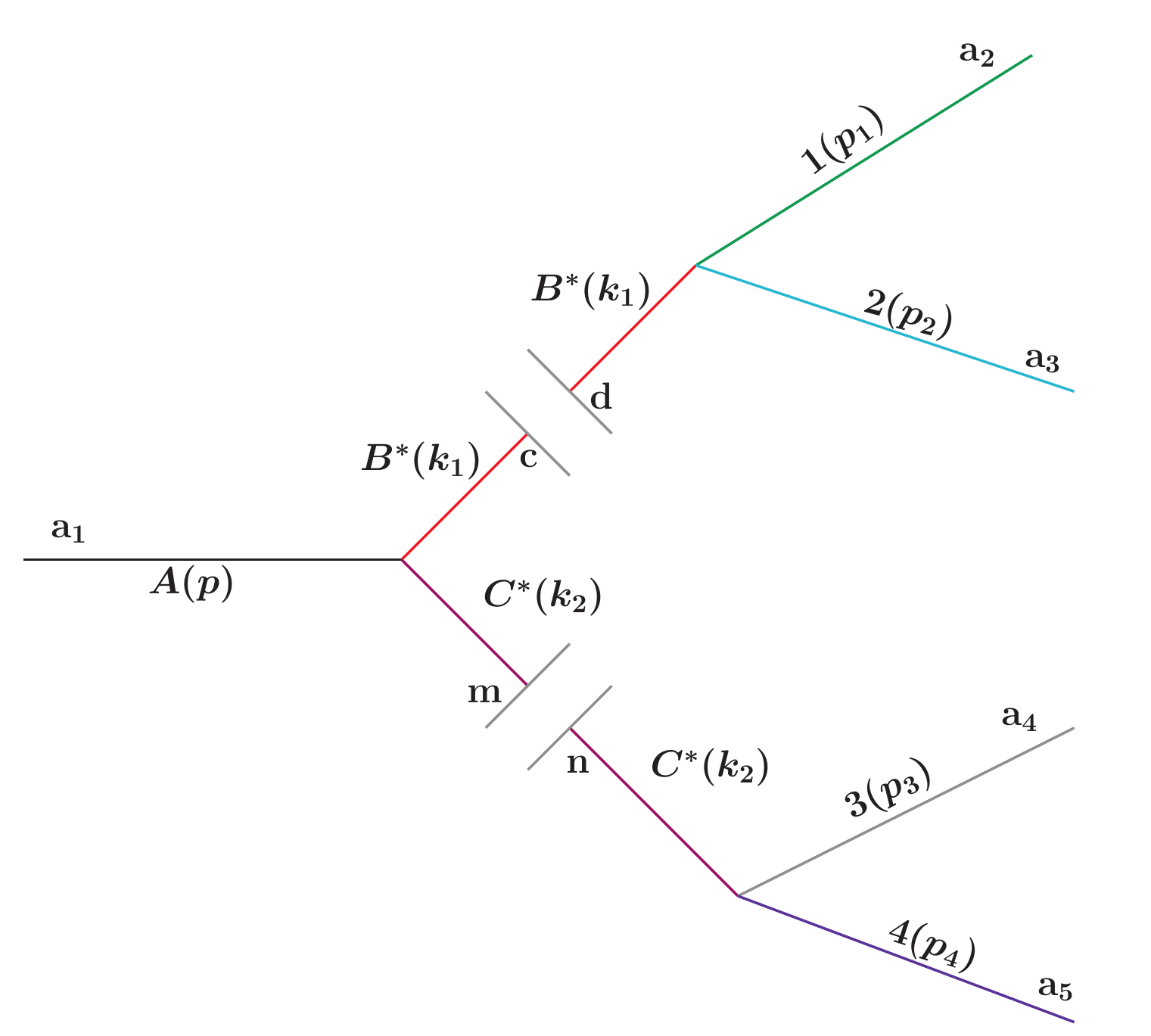,scale=0.575}
\end{center}
\caption{\label{sym_int} Sketch of the symmetric decay diagram for  $A(p)\rightarrow B^* (k_1)  C^*(k_2)\ {\rm with} \ B^*(k_1)\rightarrow 1(p_1)  2(p_2)\ \rm{and} \ C^*(k_2)\rightarrow 3(p_3)  4(p_4)$.
}
\end{figure}

\begin{figure}[h!]
\begin{center}
\epsfig{figure=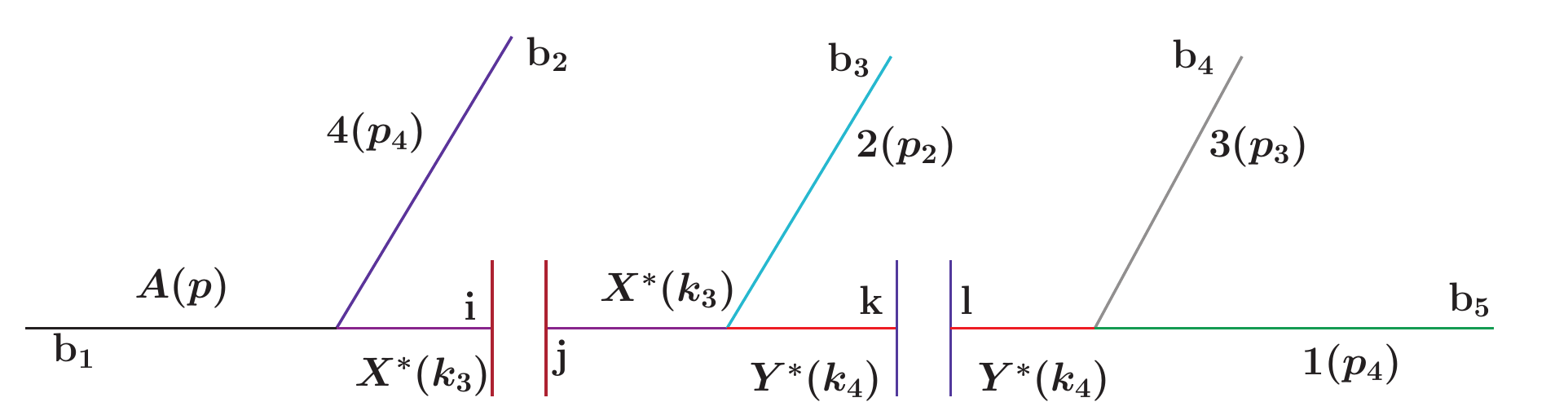,scale=0.575}
\end{center}
\caption{\label{asym_int} Sketch of the asymmetric decay diagram for  $A(p)\rightarrow 4(p_4)  X^*(k_3)\ {\rm with} \ X^*(k_3)\rightarrow 2(p_2)  Y^*(k_4)\ {\rm and} \ Y^*(k_4)\rightarrow 3(p_3)  1(p_1)$.
}
\end{figure}

Here, we are providing a brief sketch of structures of the interference term for the latter scenario. Note that unlike Figs.~(\ref{fig1.2}) and (\ref{fig2.3}), here in Figs.~(\ref{sym_int}) and (\ref{asym_int}), the off-shell propagators are different and also the positions of the external particles.

To explain this possibility let us consider the following four-body decay: $A(p)\rightarrow 1(p_1) \ 2(p_2) \ 3(p_3) \ 4(p_4)$ which can be achieved through two different cascades. One of them is  symmetric: $A(p)\rightarrow B^* (k_1) \ C^*(k_2), \ B^*(k_1)\rightarrow 1(p_1) \ 2(p_2), \ C^*(k_2)\rightarrow 3(p_3) \ 4(p_4)$, see Fig.~(\ref{sym_int}) and  the other one is asymmetric: $A(p)\rightarrow 4(p_4) \ X^*(k_3), \ X^*(k_3)\rightarrow 2(p_2) \ Y^*(k_4), \ Y^*(k_4)\rightarrow 3(p_3) \ 1(p_1)$, see Fig.~(\ref{asym_int}).
  
The interference contribution to the decay width can be given as:
\begin{align}
\Gamma_{int} (A\rightarrow 1234) \ = & \ \frac{1}{m_A}\int\left[\frac{1}{\pi} \left(\frac{d m_{12}^2}{(m_{12}^2-m_{B}^2) (m_{13}^2 - m_X^2)}\right)\right]
\int\left[\frac{1}{\pi} \left(\frac{d m_{34}^2}{(m_{34}^2-m_{C}^2) (m_{123}^2 - m_Y^2)}\right)\right] \nonumber
\\
& \ \times \frac{1}{2}\int \frac{\bar{\beta}}{8\pi}\frac{d \cos\theta}{2}\frac{d\phi}{2\pi}
\frac{1}{2}\int \frac{\bar{\beta}_{12}}{8\pi}\frac{d \cos\theta_{12}}{2}\frac{d\phi_{12}}{2\pi} \frac{1}{2}\int \frac{\bar{\beta}_{34}}{8\pi}\frac{d \cos\theta_{34}}{2}\frac{d\phi_{34}}{2\pi}
 \nonumber \\
& \ \Big[ \mathcal{M}_1(A\rightarrow B^* \ C^*) \mathcal{M}_2(B^*\rightarrow 1 \ 2) \mathcal{M}_3(C^*\rightarrow 3 \ 4)\Big]^\dagger
 \nonumber \\
& \ \Big[ \mathcal{M}_4(A\rightarrow 4 \ X^*) \mathcal{M}_5 (X^*\rightarrow 2 \ Y^*) \mathcal{M}_6(Y^*\rightarrow 3 \ 4) \Big],
\end{align}
where the necessary kinematical variables are $k_1^2 \equiv m_{12}^2 = (p_1 + p_2)^2, \ k_2^2 \equiv m_{34}^2 = (p_3 + p_4)^2, \ k_3^2 \equiv m_{123}^2 = (p_1 + p_2 + p_3)^2 $ and $k_4^2 \equiv m_{13}^2 = (p_1 + p_3)^2$. Now one can use the explicit forms of the momentum variables, like $p_1, \ p_2, \ \text{and} \ p_3$ as given in Eqns.~(\ref{sym_p_1}), (\ref{sym_p_2}) and (\ref{sym_p_3}) respectively, and  then   $m_{13}^2 \ \text{and} \ m_{123}^2$  can be written in terms of  $m_{12}^2, \ m_{34}^2 \ \text{and,} \ \theta 's \ \text{and} \ \phi 's $ to  perform phase space integration.

\section{Interaction couplings} \label{App:AppendixB}

Here, we have tabulated the interactions that we have used in our toy examples:
\begin{center}
\renewcommand{\arraystretch}{1.4}
    \begin{tabular}{ | c | c |}
   \hline
    Interactions & Couplings
     \\
     \hline
    
    $H q \bar{q}$ & $-\frac{ig m_q}{2 m_W}$   \\[1ex]
    \hline 
   $H W^{\mu+} W^{\nu-}$ & $ig m_W\eta^{\mu\nu}$   \\[1ex]
    \hline
    $W q_1 q_2$ & $\frac{ig }{2\sqrt{2}}V_{q_1 q_2}$   \\[1ex]
   \hline
       $W f_1 \bar{f}_2$ & $\frac{ig }{2\sqrt{2}}$   \\[1ex]
    \hline
       $\Phi\, q_1 \bar{q_2}$ &$-iy_{q_1q_2}$  
   \\[1ex]\hline
        $\Phi\, f_1 \bar{f}_2$ &$-iy_{f_1 f_2}$  
   \\[1ex] 
    \hline
    \end{tabular}
    \renewcommand{\arraystretch}{1.2}
\end{center}




\end{document}